\documentclass[twocolumn,amsmath,amsfonts,amssymb,aps,prd,10pt,superscriptaddress,nofootinbib,noeprint,preprintnumbers,floatfix]{revtex4-2}

\usepackage[section]{placeins}
\usepackage{slashed}

\usepackage{tikz}
\usetikzlibrary{decorations.pathreplacing,decorations.markings}
\usepackage{tikz-feynman}

\usepackage{subcaption}
\usepackage{graphicx, color}

\usepackage[letterspace=-10]{microtype} 

\usepackage{bm, amsmath, amsfonts}
\usepackage{multirow, tabularx, dcolumn}
\usepackage{mathtools} 
\usepackage{booktabs}
\usepackage{placeins}
\usepackage{soul} 

\usepackage[utf8]{inputenc} 
\usepackage{hyperref}

\usepackage{xcolor}
\definecolor{jlab_red}{RGB}{192,39,45}
\definecolor{jlab_orange}{RGB}{249,102,0}
\definecolor{jlab_blue}{RGB}{47,122,121}
\definecolor{jlab_green}{RGB}{65,125,10}
\definecolor{jlab_grey}{RGB}{125,125,125}

\hypersetup{%
pdftitle = {TITLE},
pdfsubject = {QCD},
pdfkeywords = {QCD, Hadron, Charmonium, Physics, Lattice, Meson},
pdfauthor = {Hadron Spectrum Collaboration},
colorlinks = {true},
filecolor = {black},
linkcolor = {jlab_blue},
menucolor = {black},
citecolor = {jlab_green},
urlcolor = {jlab_green},
}{}

\begin{document}

\title{Radiative decays of the $h_c$ meson from lattice QCD}

\author{Mischa~Batelaan}
\email{m.batelaan@adelaide.edu.au}
\affiliation{CSSM, Department of Physics, Adelaide University, Adelaide, 5005, South Australia, Australia}
\affiliation{Department of Physics, College of William and Mary, Williamsburg, VA 23187, USA}
\author{Jozef~J.~Dudek}
\email{jjdudek@wm.edu}
\affiliation{Department of Physics, College of William and Mary, Williamsburg, VA 23187, USA}
\author{Robert~G.~Edwards}
\email{edwards@jlab.org}
\affiliation{\lsstyle Thomas Jefferson National Accelerator Facility, 12000 Jefferson Avenue, Newport News, VA 23606, USA}

\collaboration{for the Hadron Spectrum Collaboration}
\noaffiliation

\date{\today}

\begin{abstract}
\noindent

We explore, for the first time in lattice QCD, the radiative decays to $\gamma \eta$ and $\gamma \eta'$ of the $J^{PC}=1^{+-}$ charmonium state, $h_c$. This work expands upon a previous calculation of $J/\psi$ decays to the same final states, incorporating novel technology to access the $h_c$ where it is an excited state and to separate the independent electric dipole and longitudinal form--factors. 
Results for the radiative decay rates, computed on three-flavor lattices with $m_\pi \sim $ 391 MeV show a suppression relative to experiment that is similar to that observed in the prior calculation of $J/\psi \to \gamma \eta^{(\prime)}$.
The timelike $Q^2$ dependence of both form--factors together influence the Dalitz decays, $h_c \to \ell^+ \ell^- \eta^{(\prime)}$, which are as yet unmeasured experimentally.
\end{abstract}

\preprint{JLAB-THY-26-4842}
\maketitle

\section{Introduction}\label{intro}

Radiative decays of the $J/\psi$ and other heavy-quark vector mesons copiously produced in $e^+ e^-$ collisions serve as a factory for the production of light meson resonances in a process that is theoretically relatively clean.
The BESIII experiment has collected more than ten billion $J/\psi$ and also nearly three billion of the heavier vector meson, $\psi(3686)$. Charge--conjugation positive light meson systems can be produced directly in radiative decays from these states, but also through radiative decays of an intermediate $h_c$ meson accessed through the isospin--violating $\pi^0 \, h_c$ decay of the $\psi(3686)$.

The $h_c$ is a $J^{PC} = 1^{+-}$ charmonium state having a relatively small total width under 1 MeV dominated by the radiative transition mode $h_c \to \gamma \eta_c$. Radiative decays to light hadron systems in which the $c\bar{c}$ pair annihilate contribute with branching fractions typically at the parts--per--thousand level~\cite{ParticleDataGroup:2026aaa}. The specific decays to the lightest isoscalar pseudoscalars, $h_c\to \gamma \eta$ and $h_c\to \gamma \eta'$, have been measured with good statistical precision at BES III~\cite{BESIII:2024xfs}.

\smallskip
In Refs.~\cite{Batelaan:2025vhx, Batelaan:2025vbb} we presented an application of lattice QCD to the closely related radiative decays, $J/\psi \to \gamma \eta^{(\prime)}$, which are described by a single amplitude of magnetic dipole character. We accessed the $\eta'$ as an excited state lying above the ground state $\eta$ by using \emph{optimized operators} constructed  as linear superpositions in a large basis of operators with pseudoscalar quantum numbers\footnote{following the spectrum--determination work of Ref.~\cite{Dudek:2013yja}.}. By averaging over a large number of computed three--point correlation functions, we were able to obtain statistically precise signals in a process featuring quark--antiquark annihilation. The large gap between the mass scale of charmonium and that of the light pseudoscalar mesons required us to consider large values of meson momentum in order to sample the decay kinematics corresponding to nearly real photons, $Q^2 \approx 0$. Computing for a large number of different meson momenta, we mapped out the timelike $Q^2$ dependence, which can be accessed through Dalitz decays, $J/\psi \to \ell^+ \ell^- \, \eta^{(\prime)}$, for leptons $\ell = e, \mu$~\cite{BESIII:2018qzg, BESIII:2018aao}.

This prior calculation, performed with heavier than physical light--quark masses, such that $m_\pi \sim 391 \,\mathrm{MeV}$, showed that the techniques applied are effective in determining statistically clean results, but the real photon transition rates extracted for $J/\psi \to \gamma \eta$ and $J/\psi \to \gamma \eta'$ both lie significantly below the experimental values. The origin of this discrepancy remains unknown. Could there be unexpectedly large light--quark mass dependence in this process? Could there be unusually large discretization errors that are not visible as departures from continuum--like behavior at a single lattice spacing? Might the properties of $\eta$ and $\eta'$ relevant to this process be particularly sensitive to any imperfect topological sampling of gauge fields on these particular lattices? With these questions in mind, we explore the transitions $h_c\to \gamma \eta^{(\prime)}$ on the same lattices.

The $h_c$ is the next--lightest charge--conjugation negative charmonium meson above the $J/\psi$. Within $c\bar{c}$ potential models it corresponds to the $^1\!P_1$ state, which differs in both orbital angular momentum and total heavy--quark spin from the $^3\!S_1$ $J/\psi$, and as such we might expect it to have different radiative decay systematics.
In order to study its radiative decays to $\gamma \eta^{(\prime)}$ in lattice QCD we must overcome several new challenges beyond those dealt with in Refs.~\cite{Batelaan:2025vhx, Batelaan:2025vbb}.
The reduced rotational symmetry of a cubic spatial lattice, particularly for systems with non-zero momentum, leads to irreducible representations (\emph{irreps}) which contain states having a range of $J^P$ values. In practice this means that the $h_c$ often appears together with the $J/\psi$ as an \emph{excited state} such that we cannot simply rely upon ground--state dominance at large Euclidean times -- we will use a further application of optimized operator technology to access it reliably.
The $h_c$ is somewhat heavier than the $J/\psi$ meaning that even larger light--meson momenta are required to sample the $Q^2 \approx 0$ region -- we will extend our consideration of $\eta$ and $\eta'$ mesons in--flight out to previously unexplored values.
In a transition induced by the vector current between an axial--meson and a pseudoscalar--meson, for each kinematic configuration, the transition matrix element is a linear superposition of an \emph{electric dipole} form--factor and a \emph{longitudinal} form--factor, with the weights depending upon the specific kinematics and helicities selected -- we will need to evaluate linearly--independent combinations at each photon virtuality, $Q^2$, to separate the multipoles.

\smallskip
In this paper we will present a world--first lattice QCD calculation of multipole transition form-factors for $h_c \to \gamma \eta$ and $h_c \to \gamma \eta'$ evaluated on the same $m_\pi \sim 391\, \mathrm{MeV}$ lattices used in Refs.~\cite{Batelaan:2025vhx, Batelaan:2025vbb}. The computation will make use of the technical developments presented in those papers. In addition, to maximally make use of three--point correlation functions computed for several different source--sink separations, we will introduce a procedure for simultaneous fitting of their time--dependencies.

\smallskip
The use of optimized operator technology in charmonium also allows us access to the first--excited vector state, the $\psi'$, corresponding to the experimental $\psi(3686)$ meson. We will compute transitions $\psi' \to \gamma \eta^{(\prime)}$, presenting first lattice QCD results for this process, finding a signal that is statistically compatible with zero on the current lattices.

\medskip
We refer the reader to Ref.~\cite{Batelaan:2025vbb} where they will find a description of the approach we will follow including details of the three--flavor anisotropic lattices, the construction of the improved vector current, and the method of computation of the relevant two--point and three--point correlation functions. In this paper we will mostly follow the conventions and notation introduced in that paper, and will present in detail only those technical matters which deviate from the previous method.

\section{Accessing the $h_c$ and $\eta^{(\prime)}$ at various momenta}\label{two-point}

\begin{figure}
\centerline{\includegraphics[width=.8\columnwidth]{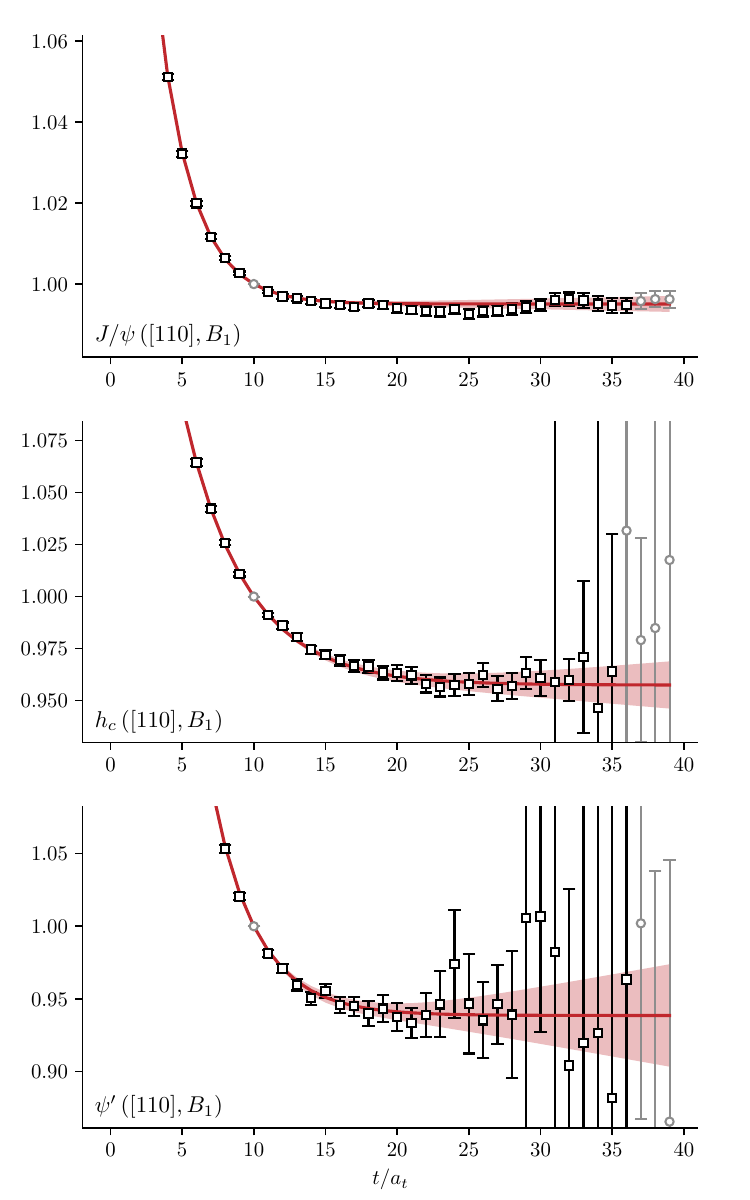}} 
\caption[]{\label{fig:prin-corr-110} 
Eigenvalues of generalized eigenvalue decomposition of matrix of correlation functions weighted by the leading time-dependence, $e^{E_{\mathfrak{n}}(t-t_{0})}\lambda_{\mathfrak{n}}(t,t_0)$, with $t_0 = 10\, a_t$, for $\mathbf{n}\, \Lambda=[110]\, B_1$. }
\end{figure}

In order to separate the electric dipole and longitudinal form--factors in the $h_c \to \gamma \eta^{(\prime)}$ process, it proves useful to consider both the helicity 0 and helicity $\pm 1$ components of the $J^{PC}=1^{+-}$ $h_c$ at many values of momentum. In--flight, the helicity $\pm 1$ components subduce into irreps which also contain $J^{PC}=1^{--}$ (see Table I of Ref.~\cite{Batelaan:2025vbb}) such that the $h_c$ is not the ground state in these cases, as it lies above the $J/\psi$. \emph{Optimized operators} following from diagonalization of a matrix of two--point correlation functions computed with a basis of fermion bilinear operators, as described in Ref.~\cite{Batelaan:2025vbb}, allow us to reliably project out the $h_c$. Figure~\ref{fig:prin-corr-110} shows the eigenvalues of a generalized eigenvalue problem for the three lightest states in the $[110]\, B_1$ irrep, which correspond to one combination of the helicity $\pm 1$ components of the $J/\psi$, the $h_c$, and the $\psi'$ mesons.
The figure shows that while the statistical quality of the $h_c$ eigenvalue is not as high as for the $J/\psi$, there is still a clear signal out to about timeslice 30. Contributions of higher excited states appearing as curvature at early times in this plot appear to be somewhat larger than for the $J/\psi$, but still modest.

\smallskip
The helicity 0 components of the $h_c$ subduce into moving--frame irreps which also contain $J^{PC}=0^{--}$ and $2^{--}$ as well as higher $J$. The $0^{--}$ case is one of \emph{exotic} quantum numbers, with the lightest such state being significantly heavier than the $h_c$. The $2^{--}$ case is non--exotic, but the lightest such state is somewhat heavier than the $h_c$, so that for the helicity 0 component, the $h_c$ is the ground--state in the irreps in which it appears. It is still accessed using an optimized operator coming from a two--point correlation function diagonalization, leading to rapid dominance of that single state with Euclidean time evolution\footnote{Appendix~\ref{app1} presents an investigation of the compatibility of form--factor determinations if one chooses not to combine data from cases where the $h_c$ is the ground--state and where it is an excited--state.}

\begin{figure}
\centerline{\includegraphics[width=.95\columnwidth]{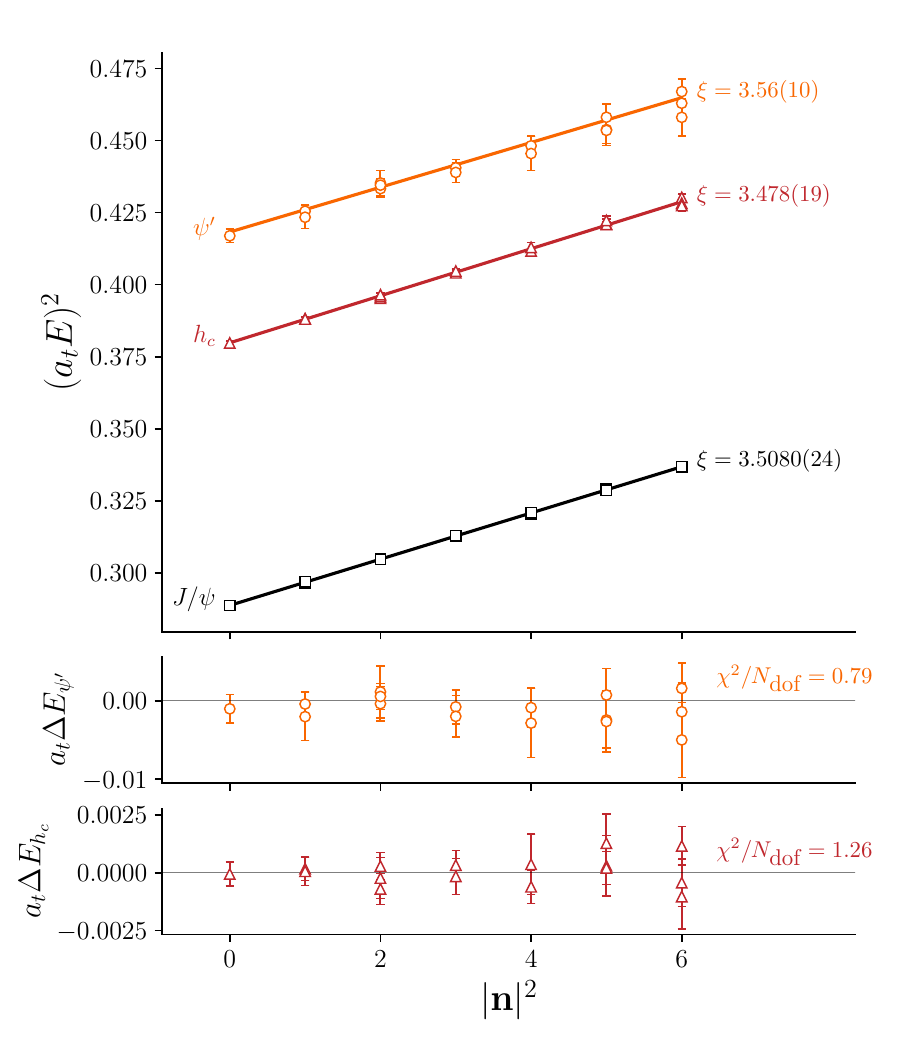}} 
\caption[]{\label{fig:ccbar_dispersion} Energies of $J/\psi$, $h_c$, and $\psi'$ mesons for momenta up to $|\mathbf{n}|^2=6$. Where a momentum has more than one energy value, these reflect the energies determined in different irreps (e.g. $A_2, B_1, B_2$ at $\mathbf{n}=[110]$ for $h_c$). Lines show continuum--like dispersion relation fits, labelled by the fitted anisotropy value, with departures from these fits shown below.}
\end{figure}

\smallskip
Two--point correlation matrix diagonalizations are performed on each relevant irrep for all momenta with $|\mathbf{n}|^2 \le 6$, and the energy as a function of momentum for the extracted $J/\psi$, $h_c$, and $\psi'$ states are shown in Figure~\ref{fig:ccbar_dispersion}. For each meson we observe no systematic departure from a continuum--like relativistic dispersion relation, and we extract broadly consistent estimates of the lattice anisotropy, $\xi$, from each.

\begin{figure}
\centerline{\includegraphics[width=.95\columnwidth]{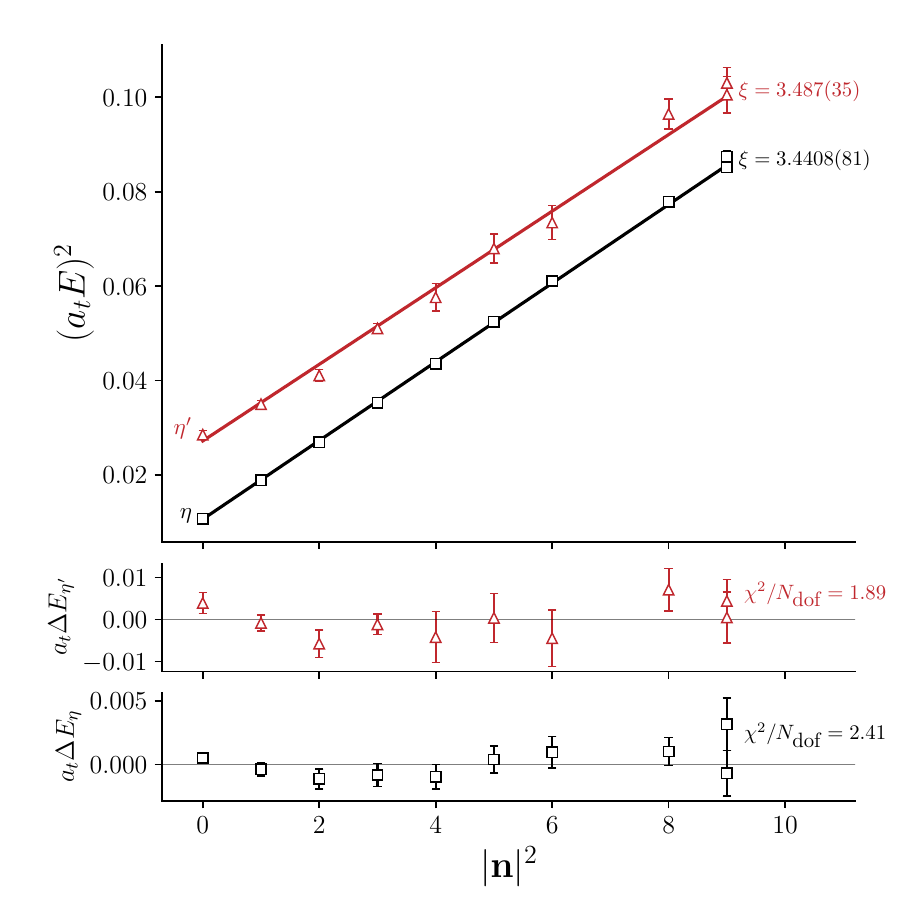}} 
\caption[]{\label{fig:eta_dispersion} Energies of $\eta$, $\eta'$ mesons for momenta up to $|\mathbf{n}|^2=9$. Lines show continuum--like dispersion relation fits, with departures from these fits shown below.}
\end{figure}

In order to access the region around $Q^2=0$ in the processes $h_c \to \gamma \eta^{(\prime)}$, owing to the heavier $h_c$, we need to consider $\eta^{(\prime)}$ with higher momentum than in the previous $J/\psi$ calculation. Figure~\ref{fig:eta_dispersion} shows the dispersion relations for the $\eta$ and $\eta'$ states for momenta out to $|\mathbf{n}|^2 = 9$. 
We observe that there is only a modest increase in statistical noise with increasing meson momentum, and no significant deviation away from the continuum--like relativistic dispersion relation. 

\section{Extraction of multipole form-factors from three-point correlation functions}\label{three-point}

The decays $h_c \to \gamma \eta^{(\prime)}$ can be characterized by a covariant vector--current matrix--element decomposition\footnote{First presented for the functionally equivalent case of $0^{+} \to 1^{-}$ in Ref.~\cite{Dudek:2006ej}. We follow the redefinition of $C_1(Q^2)$ given in Ref.~\cite{Delaney:2023fsc} which removes an explicit factor of $\sqrt{Q^2}$.},
\begin{widetext}
\begin{align}
\big\langle \eta^{(\prime)}(\mathbf{p}') \big| j^\mu(0) \big| h_c(\mathbf{p}, \lambda) \big\rangle 
= \Omega^{-1}(Q^2) \Big( &E_1(Q^2) \big[ \Omega(Q^2) \, \epsilon^\mu - (\epsilon\cdot p') \big( (p\cdot p') \,p^\mu  - m_{h_c}^2 \, p'^\mu \big) \big] \nonumber \\
&+ C_1(Q^2)\, m_{h_c} (\epsilon\cdot p') \big[  (p\cdot p') \, (p+p')^\mu - m_{\eta^{(\prime)}}^2 \,p^\mu  - m_{h_c}^2 \,p'^\mu \big]     \Big) \, , \label{eqn:decomp}
\end{align}
\end{widetext}
where $\epsilon^\mu(\mathbf{p}, \lambda)$ is the covariant polarization vector for the definite--helicity $h_c$ state, and where the kinematic function, $\Omega(Q^2) = (p \cdot p')^2 - m_{h_c}^2 m_{\eta^{(\prime)}}^2$. 

\pagebreak
This decomposition implements conservation of the vector current, $0 = (p-p')_\mu \, \big\langle \eta^{(\prime)}(\mathbf{p}') \big| j^\mu(0) \big| h_c(\mathbf{p}, \lambda) \big\rangle$, by construction. The invariant form--factors, $E_1(Q^2)$ and $C_1(Q^2)$, represent multipoles of electric dipole and `longitudinal' nature respectively~\cite{Durand:1962zza}. When considered in the rest--frame of the $h_c$, the resulting set of non--zero helicity amplitudes for decay to a (virtual) photon moving along the $z$-axis, $h_c(\mathbf{0}, \lambda) \to \gamma( q\hat{\mathbf{z}}, \lambda_\gamma) \, \eta^{(\prime)}(- q\hat{\mathbf{z}})$, are 
\begin{align*}
 A_{\lambda = + 1, \lambda_\gamma = + 1}  &= A_{\lambda = - 1, \lambda_\gamma = - 1} = - E_1(Q^2), \\
 A_{\lambda = 0, \lambda_\gamma = 0} &= - \sqrt{-Q^2} \cdot C_1(Q^2) \, ,
\end{align*}
showing the transverse nature of the electric dipole form--factor and that the contribution of the longitudinal form--factor vanishes for the real photon case ($Q^2=0$). 

\smallskip
The $h_c$ radiative decay rate (for an on--shell photon) can be obtained by explicitly including the $\frac{2}{3} e$ charge factor relevant for the photon coupling to the charm--quark, 
\begin{equation}
\Gamma(h_c \to \gamma \eta^{(\prime)}) = \frac{4}{27} \alpha \frac{|\mathbf{q}|}{m_{h_c}^2} \big| E_1(0) \big|^2 \, ,
\label{eqn:rad_decay}
\end{equation}
where $|\mathbf{q}|$ is the magnitude of the $\eta^{(\prime)}$ or photon momentum in the rest frame of the $h_c$.

\smallskip
The form-factors $E_1(Q^2)$ and $C_1(Q^2)$ \emph{both} play a role in describing Dalitz decays, $h_c \to \ell^+ \ell^- \, \eta^{(\prime)}$, for timelike $Q^2 < - (2 m_\ell)^2$ where the virtual photon can produce a lepton--antilepton pair. In terms of $q^2 \equiv - Q^2$, the differential decay rate for this process can be expressed as,
\begin{widetext}
\begin{equation}
\frac{d\Gamma}{dq^2\, d \cos \theta_\ell^\star} 
= \frac{1}{27} \alpha^2 \frac{|\mathbf{q}|}{m_{h_c}^2}\, \frac{|\mathbf{p}_\ell^\star|}{\sqrt{q^2}}
\, \frac{1}{q^2}
\left[ \Big( \big(1+ \tfrac{4m_\ell^2}{q^2}\big) |E_1|^2 + q^2 |C_1|^2 \Big) 
+ 4 \frac{|\mathbf{p}_\ell^\star|^2}{q^2} \Big( |E_1|^2 - q^2 |C_1|^2 \Big)\, \cos^2\theta_\ell^\star
\right]\, , \label{eqn:dalitz}
\end{equation}
\end{widetext}
where $|\mathbf{q}|$ represents the magnitude of momentum of the $\eta^{(\prime)}$ in the rest--frame of the decaying $h_c$, and $(|\mathbf{p}_\ell^\star|, \theta_\ell^\star)$ represent the magnitude and direction of momentum of the $\ell^+$ in the rest--frame of the $\ell^+ \ell^-$ system, relative to the axis defined by the momentum of the $\ell^+ \ell^-$ pair in the $h_c$ rest--frame\footnote{This equation, when converted to use the different decomposition adopted in Ref.~\cite{Becirevic:2026gaf}, agrees with that presented therein up to a factor of 4, reflecting the coupling to both $c$ and $\bar{c}$ in a charmonium \emph{transition}. }.
The total width for the Dalitz decay follows from the (straightforward) integration over $\theta_\ell^\star$, and the integration over $q^2$ from $(2m_\ell)^2$ to $(m_{h_c} - m_{\eta^{(\prime)}})^2$, which requires us to have determined the $Q^2$ dependence of the form--factors $E_1(Q^2)$ and $C_1(Q^2)$ over that range.

\begin{figure*}
  \centerline{\includegraphics[width=\textwidth]
    {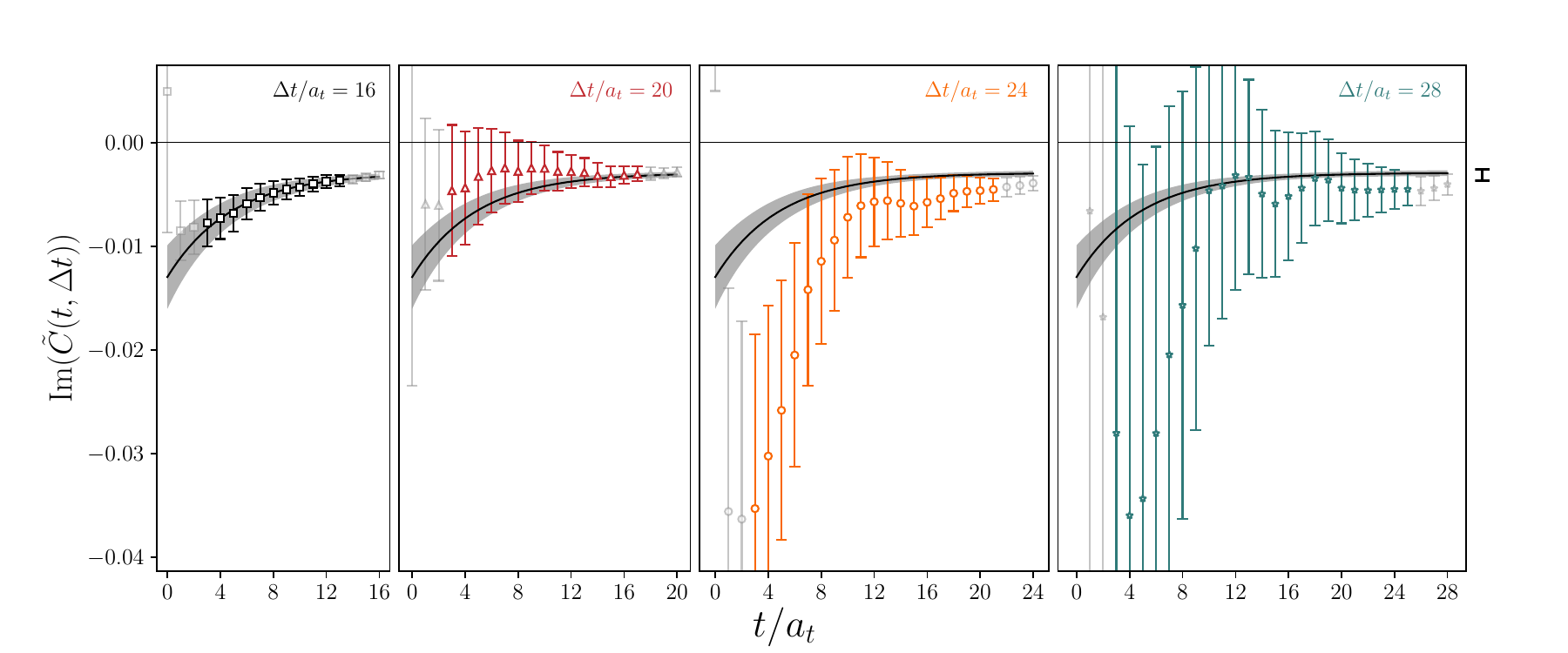}
  }
\caption[]{\label{fig:proj0-dt-comp1} An example of the result of simultaneous $\Delta t$ fitting of $\tilde{C}(t, \Delta t)$ for one particular correlation function constructed to describe $h_c \to\gamma\, \eta$. The sink and source operators have momenta $\mathbf{n}_{\eta} = [\text{-}1,\text{-}1, \text{-}2]$, $\mathbf{n}_{h_c} = [0,1,0]$ respectively, so that the virtuality of the photon is $Q^2 = -2.21 \, \mathrm{GeV}^2$. The band shows the highest AIC--value combined $\Delta t$ fit, and the bold data--point on the far right shows the AIC--weighted average value of $J$.
}
\end{figure*}

\bigskip
In order to obtain $E_1$ and $C_1$ within lattice QCD, we nonperturbatively evaluate the three--point correlation functions, 
\begin{equation*}
\label{eq:13}
  C(t, \Delta t) =  \big\langle 0 \big|  \Omega_{\eta^{(\prime)}}(\Delta t) \; j(t) \; \Omega^{\dagger}_{h_c}(0) \big| 0 \big\rangle \, ,
\end{equation*}
for all current insertion times, $t$, between the fixed source--sink separation, $\Delta t$. The use of optimized operators to interpolate  the $h_c$ and $\eta^{(\prime)}$ leads us to expect dominance of those states in the time--dependence such that, 
\begin{align}
  \tilde{C}(t, \Delta t) &\equiv \sqrt{2 E_{\eta^{(\prime)}} \, 2 E_{h_c}} \; e^{E_{\eta^{(\prime)}} (\Delta t - t)} e^{E_{h_c} t}  \,  C(t, \Delta t) \nonumber \\
  &= \langle \eta^{(\prime)} | j(0) | h_c \rangle + \ldots \, , \label{eqn:Ctilde}
\end{align}
where the ellipsis represents any residual time--dependence due to other states in the spectrum.

Such correlation functions can be computed for a large number of momentum choices for the $h_c$, the $\eta^{(\prime)}$, and for the current, allowing access to a range of $Q^2$ values. All operators lie in the basis of \emph{irreps} of the boosted lattice symmetry group, so the extracted matrix--element corresponds to a \emph{subduction} of the decomposition in Eqn.~\ref{eqn:decomp}. The practical challenge, then, is that for generic kinematical combinations accessible in the lattice calculation, this matrix--element at each discrete $Q^2$ is a linear combination of unknowns, $E_1(Q^2)$ and $C_1(Q^2)$. We generalize the expression given in Ref.~\cite{Batelaan:2025vbb} in the case of a single form--factor, to,
\begin{align*}
  \begin{split}
  &\langle \eta^{(\prime)}_{\Lambda'}(\mathbf{p}') | j^{\mathbf{q}\Lambda_{\gamma} \mu_{\gamma}} | {h_c}_{,\Lambda \mu}(\mathbf{p})\rangle \\
    &\quad = \sum\nolimits_{i} K_i(\mathbf{p'}\Lambda';\, \mathbf{q}\Lambda_\gamma \mu_\gamma; \, \mathbf{p}\Lambda\mu) \cdot  F_i(
Q^2)  \, ,
\end{split}
\end{align*}
where the sum runs over $E_1$ and $C_1$ with each having a kinematic pre--factor given by the appropriate subduction of Eqn.~\ref{eqn:decomp}.

As we did in Ref.~\cite{Batelaan:2025vbb}, we proceed under the assumption of at--most modest discretization effects, so that the Lorentz decomposition in Eqn.~\ref{eqn:decomp} relates, to a good approximation, different irreps under subduction. With a sufficient set of momentum directions, irreps, and rows, all at the same value of $Q^2$, we can form an overconstrained linear system, $\mathbf{\Gamma} = \mathbf{K}\cdot \mathbf{F}$, whose solution we take to be
\begin{equation}
\mathbf{F} = \big( \mathbf{K}^T \mathbf{\Sigma}^{-1} \mathbf{K} \big)^{-1}\, \mathbf{K}^T  \mathbf{\Sigma}^{-1} \, \mathbf{\Gamma}\, , \label{eqn:SVD}
\end{equation}
where the vector $\mathbf{\Gamma}$ holds the ($N$) computed matrix--element values, $\mathbf{K}$ is the ($N\times 2$ rectangular) matrix of kinematic factors, and $\mathbf{F}$ is the two--component vector holding the values of $E_1$ and $C_1$. We choose to introduce $\mathbf{\Sigma}$, the ($N \times N$) data covariance matrix for the $\mathbf{\Gamma}$ values, to more accurately account for correlations.
In those cases where the kinematic factors are purely real or purely imaginary we discard the part of the matrix--element which contains no signal -- this is the case for a large majority of the kinematic combinations we consider. In all cases we have a system which is constrained ($N=2$) or overconstrained ($N>2$). We perform the solution under jackknife, resetting small singular values in the computation of $\mathbf{\Sigma}^{-1}$ if required.

\medskip
In order to obtain the matrix--element values that go into $\mathbf{\Gamma}$, we have to fit the time--dependence in Eqn.~\ref{eqn:Ctilde}, accounting for any residual contributions from excited states that are not completely eliminated by the use of optimized operators.
In the calculation of $J/\psi \to \gamma \eta^{(\prime)}$ presented in Refs.~\cite{Batelaan:2025vhx, Batelaan:2025vbb}, we computed three--point correlators for four different source--sink time--separations, $\Delta t/a_t = \{12, 16, 20, 24\}$, and chose to perform \emph{independent} time--dependence fits on each using a form allowing for excited--state pollution at source and/or sink,
\begin{equation}
\tilde{C}(t, \Delta t) = J + A_{\text{src}}\,  e^{-\delta E_{\text{src}} t}   + A_{\text{snk}} \, e^{-\delta E_{\text{snk}} ( \Delta t - t )} \, . \label{eqn:tslice_fit}
\end{equation}
We found the results of these fits to different $\Delta t$ to be broadly consistent, and opted to use the $\Delta t/a_t = 16$ results as representative.

In the current calculation we again compute on multiple time--separations, in this case $\Delta t/a_t = \{16, 20, 24, 28\}$, but we now introduce a procedure of \emph{simultaneous} fitting.
Allowing for all possible time--windows in each $\Delta t$ would lead to an unmanageable number of fits, and as such we adopt a simplified scheme where, beginning with the complete set of timeslices, we systematically remove the same number of points from the source and/or the sink on every $\Delta t$ simultaneously, while ensuring that at least two timeslices are retained on the smallest $\Delta t$. Variations of Eqn.~\ref{eqn:tslice_fit} are considered with just a constant, a constant plus a source exponential, a constant plus a sink exponential, or a constant with both a source and sink exponential. The fit parameters, a subset of $\{ J,A_{\text{src}},  \delta E_{\text{src}}, A_{\text{snk}}, \delta E_{\text{snk}} \}$, are chosen to be common for all $\Delta t$. Each time--window and parameterization variation is considered and the fit characterized by the value of the Akaike Information Criterion (AIC), with the 30 largest values retained and used in a model average~\cite{Jay:2020jkz} to yield a single value of $J$ with, we argue, a conservatively estimated uncertainty.

Figure~\ref{fig:proj0-dt-comp1} presents an example of a fit description that can be obtained through this procedure. The curves show the single description having the highest value of AIC which in this case happens to have an excited state contribution only at the source. 

More details of the fitting procedure are provided in Appendix~\ref{app4} where comparison with independent fitting of each $\Delta t$ is also illustrated.

\bigskip
Our procedure was to compute every three--point correlation function capable of yielding a $h_c \to \gamma \eta^{(\prime)}$ matrix--element using an optimized $h_c$ operator with $|\mathbf{n}_{h_c}|^2 \le 2$, an optimized $\eta$ or $\eta'$ operator with $|\mathbf{n}_{\eta^{(\prime)}}|^2 \le 9$, and a vector--current insertion with $|\mathbf{n}_\mathbf{q}|^2 \le 16$. As in Ref.~\cite{Batelaan:2025vbb}, only a single momentum direction  was used for each magnitude of $\mathbf{q}$, but all directions of $\mathbf{p}$ and $\mathbf{p}'$ were computed, as well as all rows $\mu$ of each irrep $\Lambda$. Correlation functions that are related by allowed lattice rotations were averaged using the Wigner--Eckart approach presented in Section V.A of Ref.~\cite{Batelaan:2025vbb}. This was repeated for each of $\Delta t / a_t = \{ 16, 20, 24, 28 \}$, and the four resulting averaged correlation functions having the same kinematic configuration were fitted together using the procedure described above to yield a single (in general complex) value of matrix--element, $J$, to go into $\mathbf{\Gamma}$. All kinematic configurations having the same $Q^2$ value were included together in each instance of $\mathbf{\Gamma}$, and the (over)constrained linear system presented above was solved to give (real) values of $E_1$ and $C_1$ at that $Q^2$ value. This was repeated independently for each accessible value of $Q^2$ to deliver discretely sampled $E_1(Q^2)$ and $C_1(Q^2)$.

\begin{figure}
  \centerline{\includegraphics[width=.95\columnwidth]{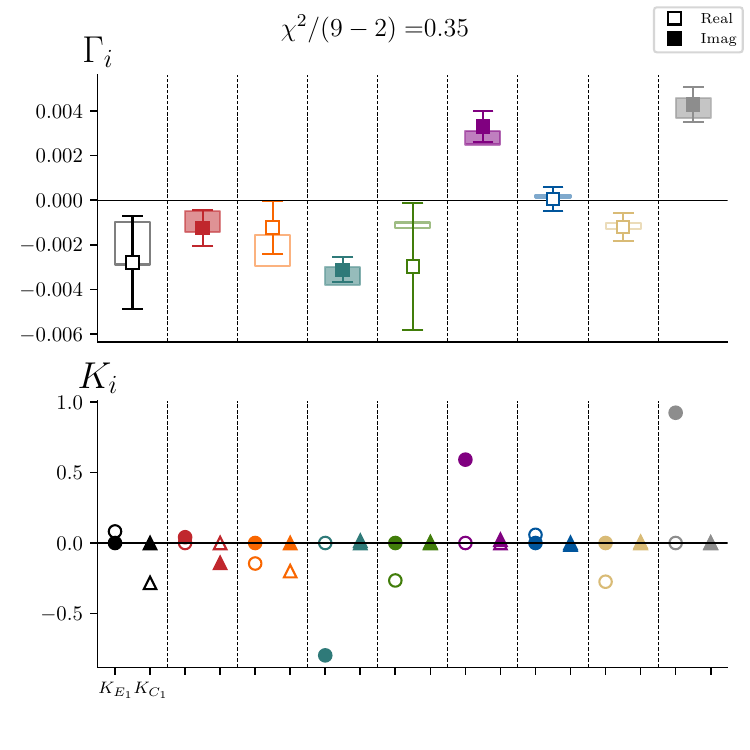}}
\caption[]{\label{fig:SVD-ex1} The upper panel shows the real (open symbols) and imaginary (filled symbols) parts of the matrix--elements from the fits to each correlator. The rectangular bands show the reconstructed matrix--elements using the extracted form--factor values.
  The lower panel shows the real and imaginary parts of the two kinematic factors for each correlator.
  The matrix--element in the fourth column is the one extracted from the fit to the timeslice data shown in Figure \ref{fig:proj0-dt-comp1}.
}
\end{figure}

\smallskip
Figure~\ref{fig:SVD-ex1} illustrates the description of  $\mathbf{\Gamma}$ in terms of just the two real numbers, $\{ E_1, C_1 \}$, at a particular $Q^2$ point. In the upper panel, the agreement between the form--factor model description and the input matrix--element data can be clearly seen, while the lower panel shows how the contributions of $E_1$ and $C_1$ to each $\Gamma_i$ vary through the differing kinematic factors. The description can be characterized by a correlated $\chi^2$ value which is minimized by the solution in Eqn.~\ref{eqn:SVD}. We choose to retain only those $Q^2$ points where $\chi_{\Gamma}^2/(N-2) < 2.0$ for further analysis (see Appendix~\ref{app2})\footnote{There are some instances in which the system is only minimally constrained, $N=2$, where we cannot define this $\chi^2$ per d.o.f. We choose to retain these points which generally match the trend of the rest of the data points obtained from $N > 2$. }.

\medskip
For $h_c\to \gamma \eta$ we generated 2298 correlation functions at each $\Delta t$, which reduced to 417 independent cases under Wigner--Eckart averaging, and these then had their timeslice dependence fitted. The resulting matrix--element values sample 76 discrete $Q^2$ values, with 53 of these surviving our cut on $\chi^2_\Gamma$.
For $h_c\to \gamma \eta'$, a similar procedure leads to estimates of $E_1(Q^2)$ and $C_1(Q^2)$ at 63 points.

\section{Form--factors}\label{form_factors}

\begin{figure*}[h]
  \centerline{\includegraphics[width=.95\textwidth]{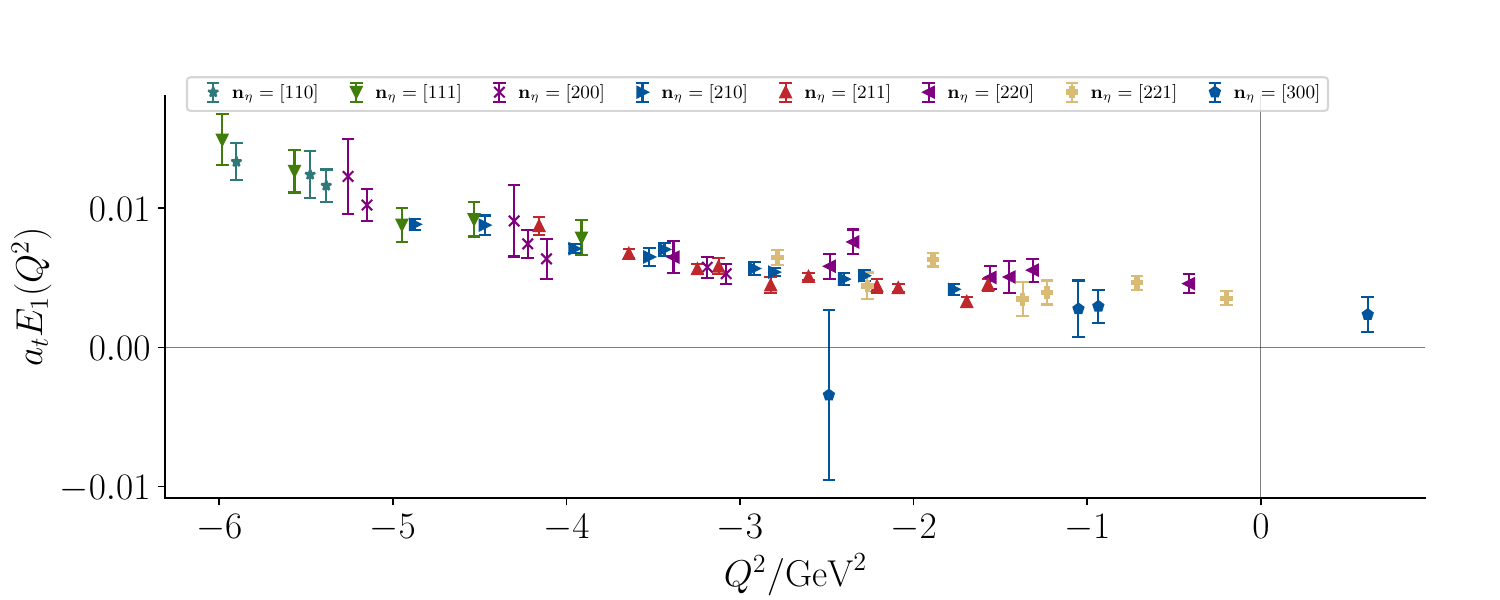}}
  \centerline{\includegraphics[width=.95\textwidth]{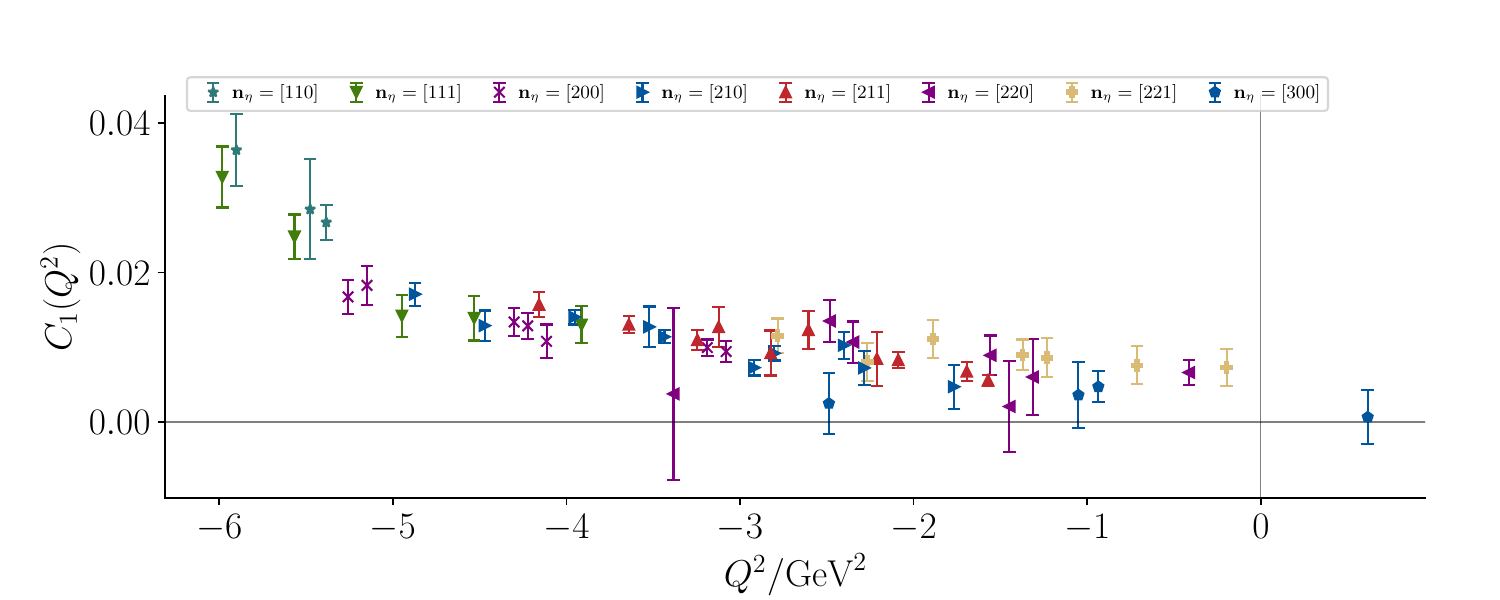}}
\caption[]{\label{fig:eta_data} Electric dipole (top) and longitudinal (bottom) form--factors for $h_c \to \gamma \eta$ as a function of photon virtuality, $Q^2$. }
\end{figure*}

The procedure described above leads to a set of values of $E_1(Q^2)$ and $C_1(Q^2)$ discretely sampled in $Q^2$ for both $h_c \to \gamma \eta$ and $h_c \to \gamma \eta'$. 
In Ref.~\cite{Batelaan:2025vbb} we described the use of explicit parameterizations to capture the $Q^2$ dependence of the single form--factor for each of $J/\psi \to \gamma \eta$ and $J/\psi \to \gamma \eta'$. We presented fits using a dipole form and two exponential--based forms, as well as a rather flexible approach based upon a conformal mapping of $Q^2$ into a variable, $z$, in terms of which a low--order polynomial description could be found. Evaluating these various fitted parameterizations at $Q^2 = 0$ yielded a set of estimates of the real photon transition. In addition, we also considered a simple linear interpolation of a few discrete points straddling $Q^2=0$ to obtain another estimate of $F(0)$.

In the current case we again find data points consistent with monotonic behavior across the explored $Q^2$ region for both $E_1(Q^2)$ and $C_1(Q^2)$, and we consider the same fit--forms\footnote{Details are presented in Appendix~\ref{app3}.}.

\subsection{$h_c \to \gamma \eta$}

Figure~\ref{fig:eta_data} shows the discrete values of $E_1(Q^2)$ and $C_1(Q^2)$ for $h_c \to \gamma \eta$ extracted from three--point correlation functions as described above. The data points obtained at different $\eta$ momenta are observed to be in excellent agreement in the regions in which they overlap, and we see clearly our reasons for going to large momentum values for the $\eta$, as the region around $Q^2=0$ is constrained only by $\mathbf{n}_\eta = [220], \, [221], \, [300]$.

For the most--timelike values of $Q^2$ considered, ${Q^2 \lesssim -6.2 \,\textrm{GeV}^2}$, which mainly come from the lowest values of $\eta$ momentum, we were not able to find reliable timeslice fits to the rescaled correlators, $\tilde{C}(t,\Delta t)$, which do not show even approximately flat behavior in $t$. As such we have excluded these points from the plots and from further analysis.

\begin{figure}[h!]
  \centerline{\includegraphics[width=.5\textwidth]{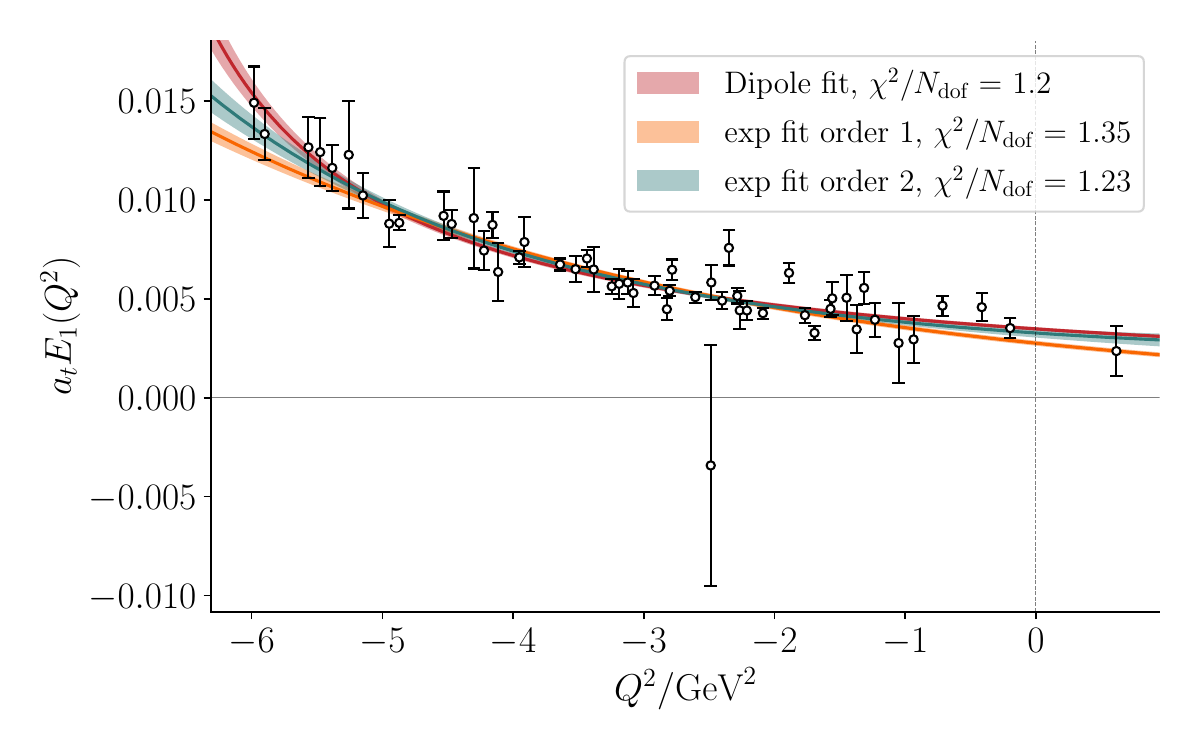}}
  \centerline{\includegraphics[width=.5\textwidth]{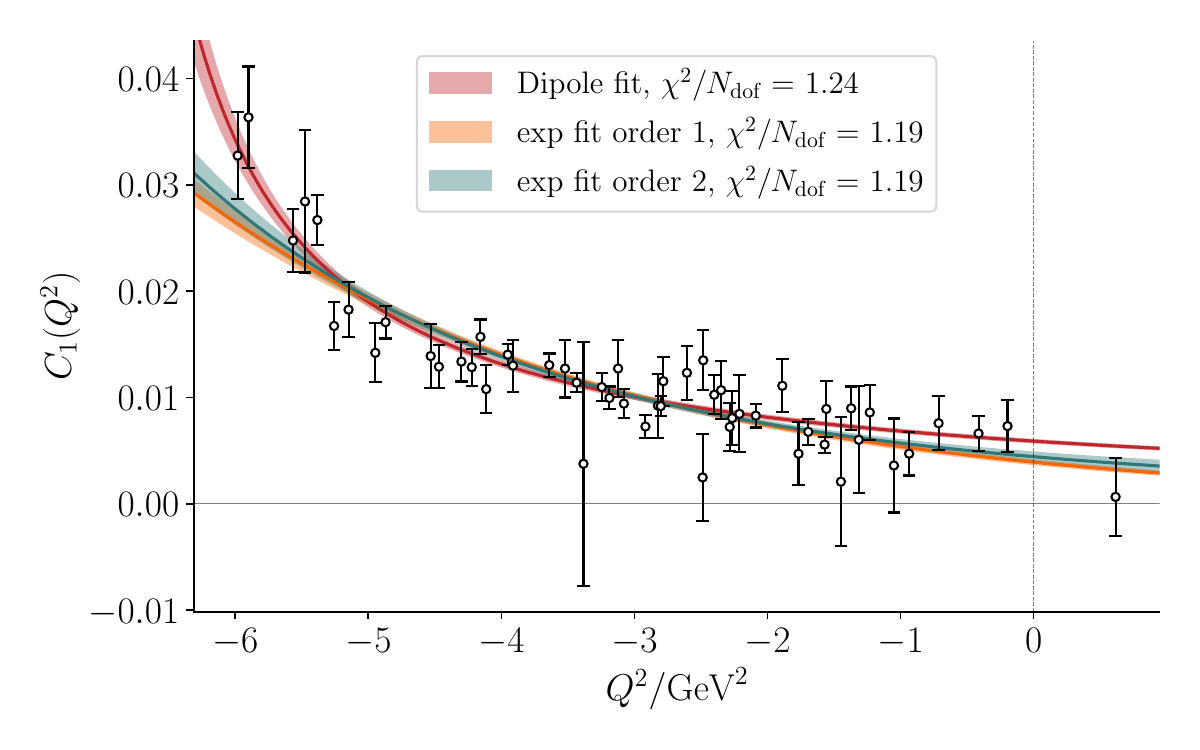}} 
\caption[]{\label{fig:eta_Qsq_dipole_exp} Electric dipole (top) and longitudinal (bottom) form--factors for $h_c \to \gamma \eta$ described by dipole and exponential fits. }
\end{figure}

\pagebreak
Figure~\ref{fig:eta_Qsq_dipole_exp} shows descriptions of the extracted form--factor data sets for $E_1(Q^2)$ and $C_1(Q^2)$ in terms of dipole and exponential forms, which are seen to agree closely with each other and the data except at the edges of the data set, $Q^2 \lesssim -5 \, \mathrm{GeV}^2$ and $Q^2 > 0$.

\begin{figure}
  \centerline{\includegraphics[width=.5\textwidth]{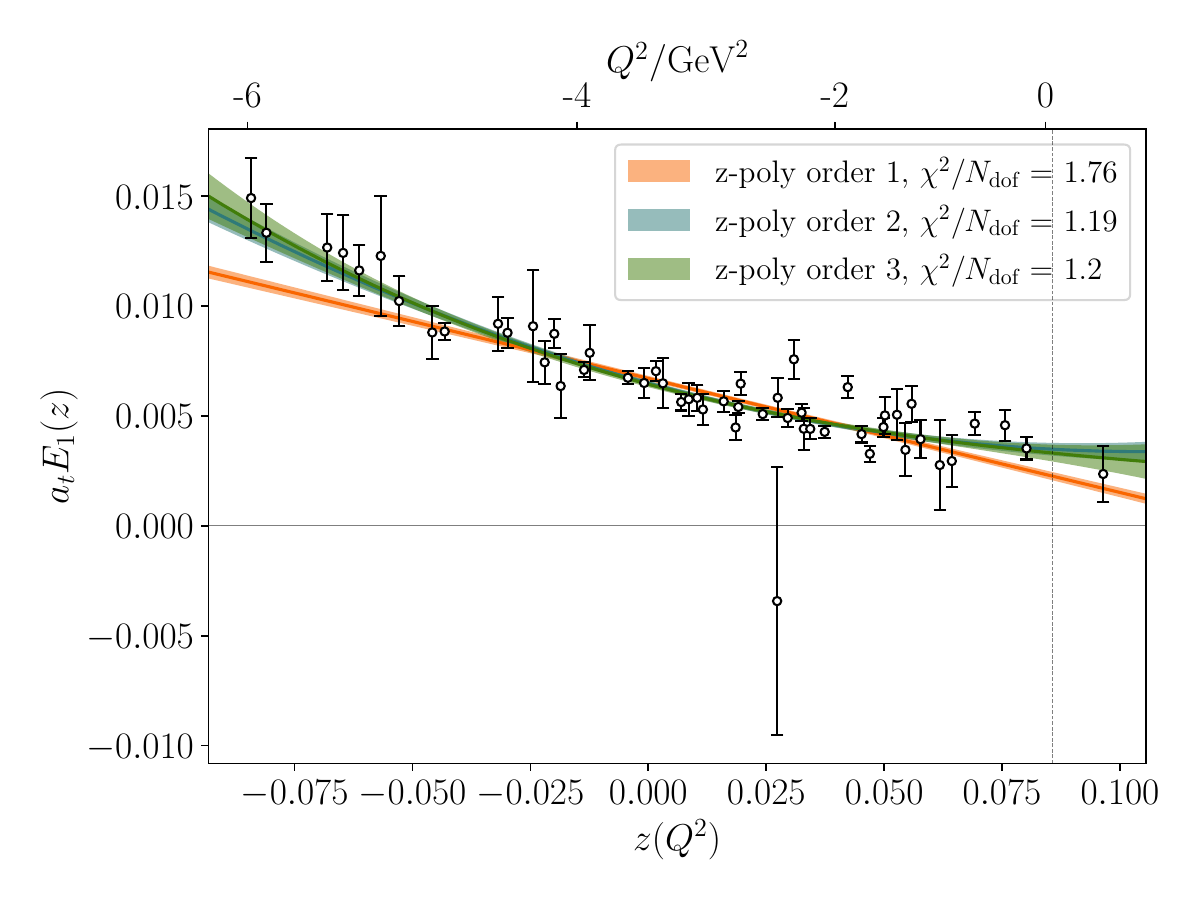}}
  \centerline{\includegraphics[width=.5\textwidth]{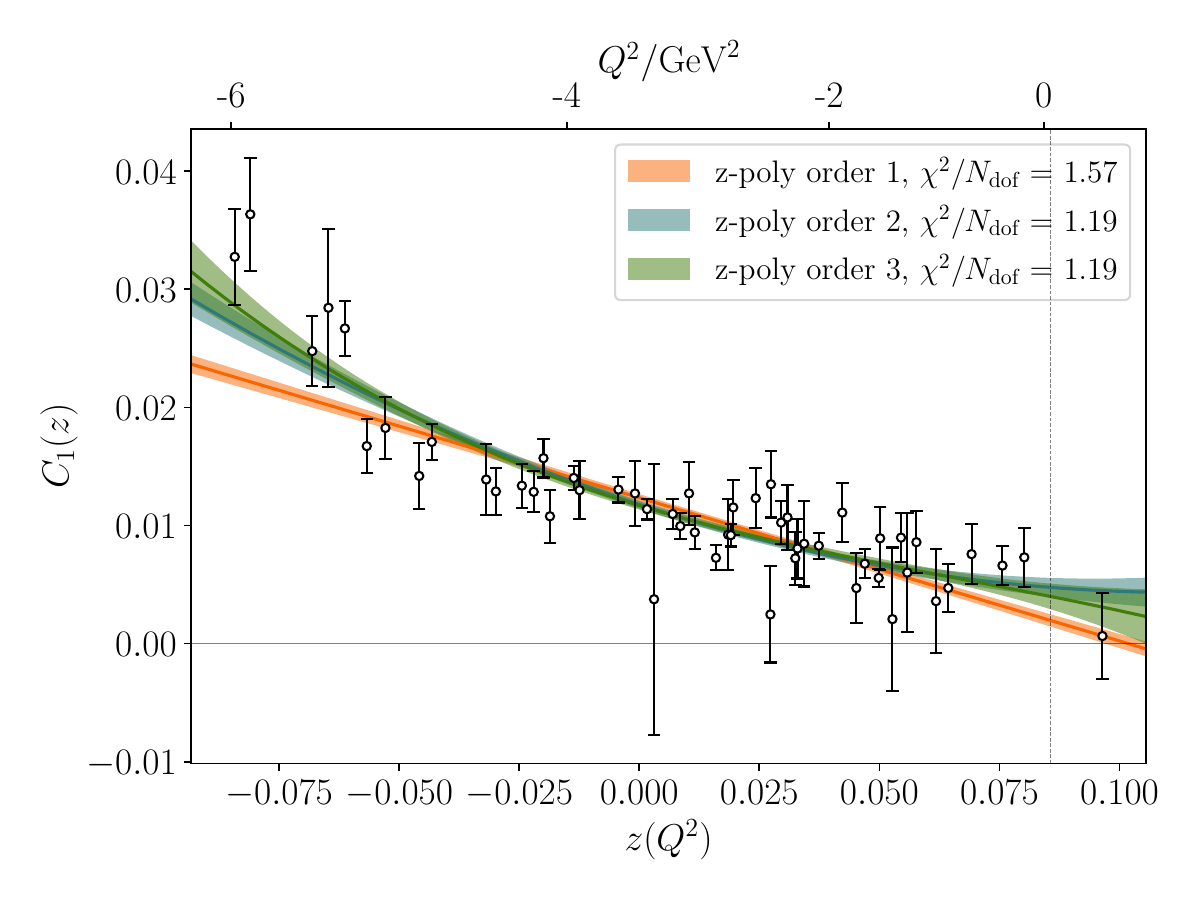}} 
\caption[]{\label{fig:eta_Qsq_z_fits} Electric dipole (top) and longitudinal (bottom) form--factors for $h_c \to \gamma \eta$ described using low--order polynomials in a conformally mapped variable $z(Q^2)$. }
\end{figure}

Figure~\ref{fig:eta_Qsq_z_fits} shows descriptions using low--order polynomials in a conformally mapped variable, $z(Q^2)$, as described in Ref.~\cite{Batelaan:2025vbb} and in Appendix~\ref{app3}. The quality of fit appears to suggest that a polynomial in $z$ of at least order 2 is required to describe the data over the wide $Q^2$ range considered.

We observe from the fit curves that the paucity of discrete points around $Q^2=0$ does lead to a degree of parameterization dependence in estimates of $E_1(0)$, but one that is still modest. The fit parameter values for all presented fits are provided in Appendix~\ref{app3}.

\pagebreak
\subsection{$h_c \to \gamma \eta'$}

\begin{figure*}
  \centerline{\includegraphics[width=.9\textwidth]{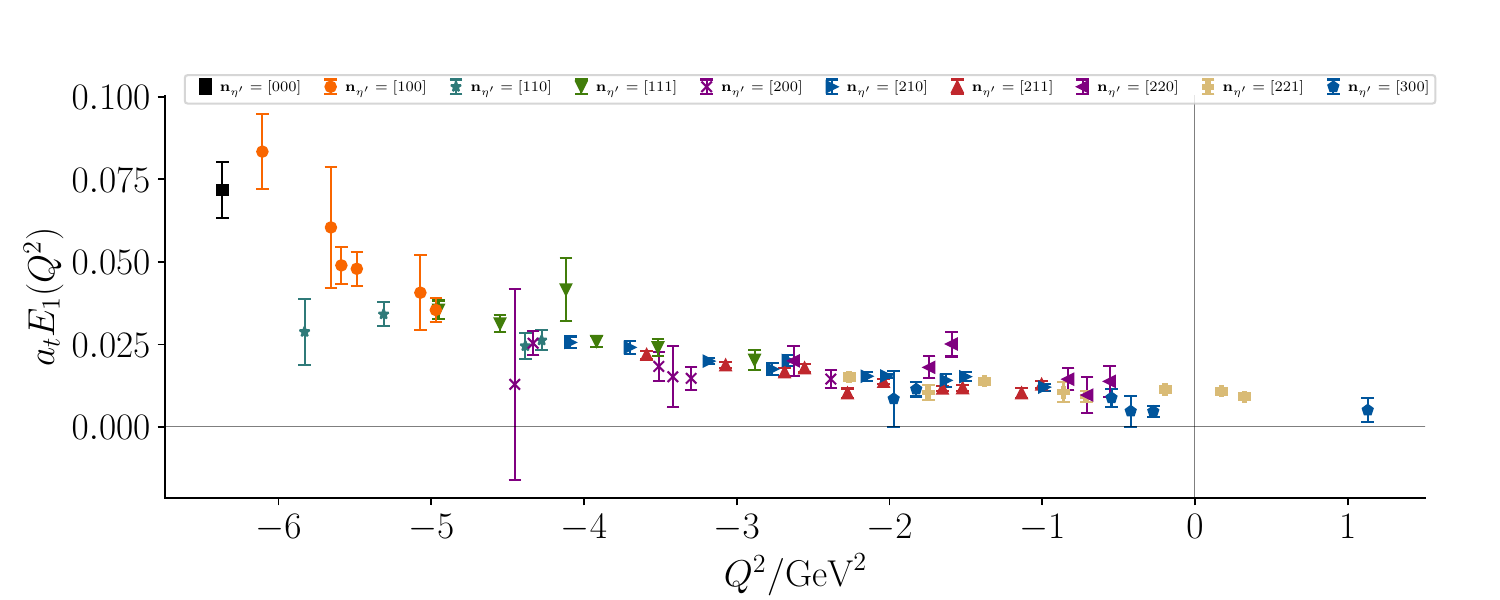}}
  \centerline{\includegraphics[width=.9\textwidth]{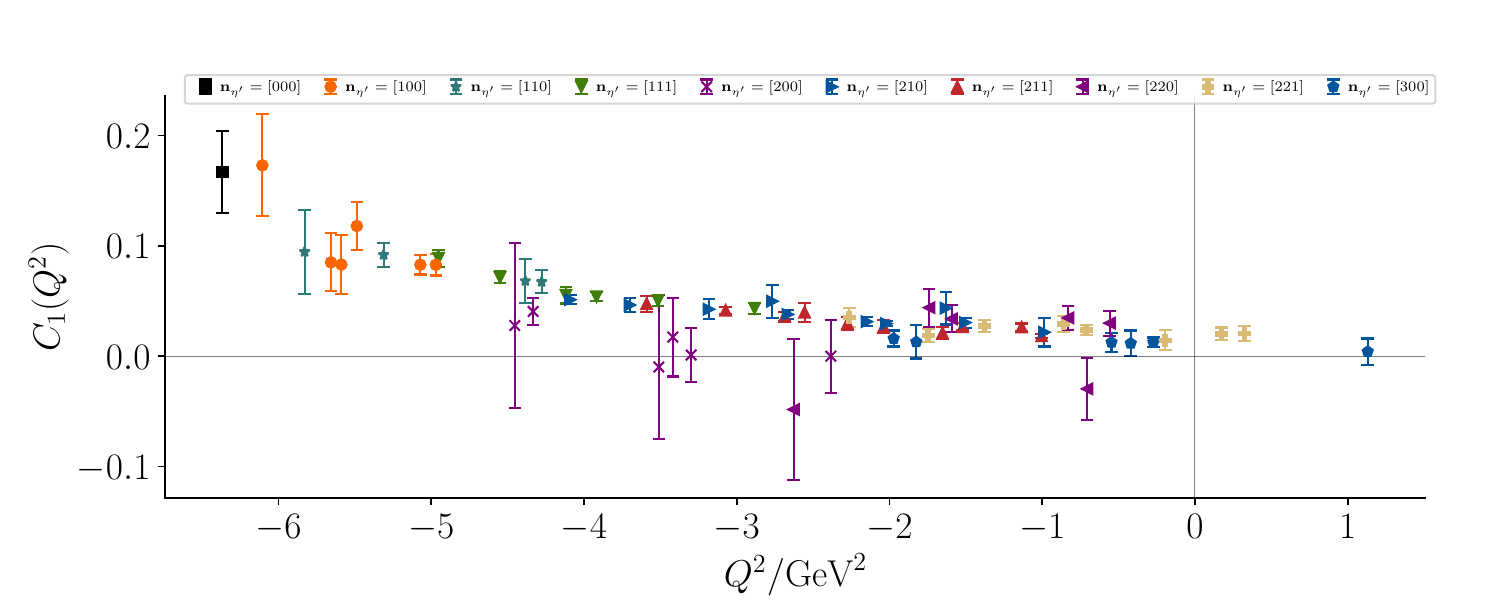}}
\caption[]{\label{fig:eta-prime_data} Electric dipole (top) and longitudinal (bottom) form--factors for $h_c \to \gamma \eta'$ as a function of photon virtuality, $Q^2$. }
\end{figure*}

Figure~\ref{fig:eta-prime_data} shows the discrete values of $E_1(Q^2)$ and $C_1(Q^2)$ for $h_c \to \gamma \eta'$ extracted as described above. As for the $\eta$ we observe generally good agreement between extractions from different $\eta'$ momenta in the regions where they overlap. Similar to the case of the radiative decays of the $J/\psi$ presented in Ref.~\cite{Batelaan:2025vbb}, we observe a somewhat larger magnitude for decays of the $h_c$ to $\eta'$ relative to $\eta$.

\begin{figure}
  \centerline{\includegraphics[width=.5\textwidth]{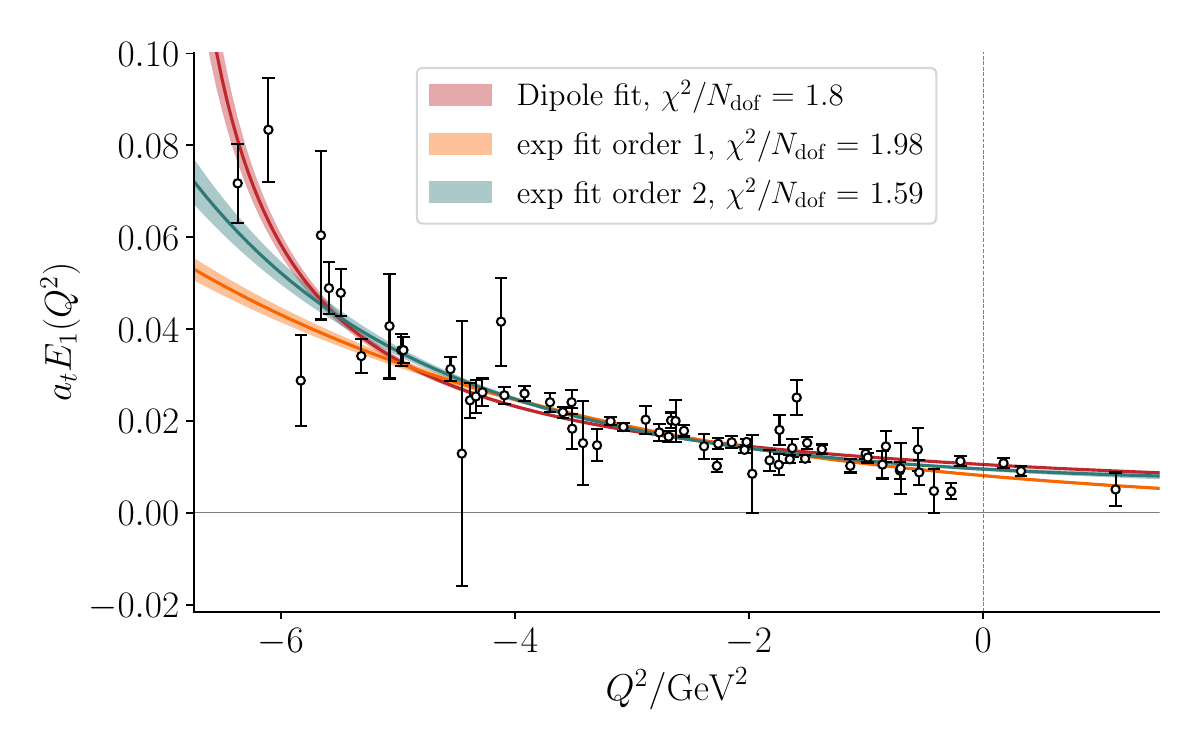}}
  \centerline{\includegraphics[width=.5\textwidth]{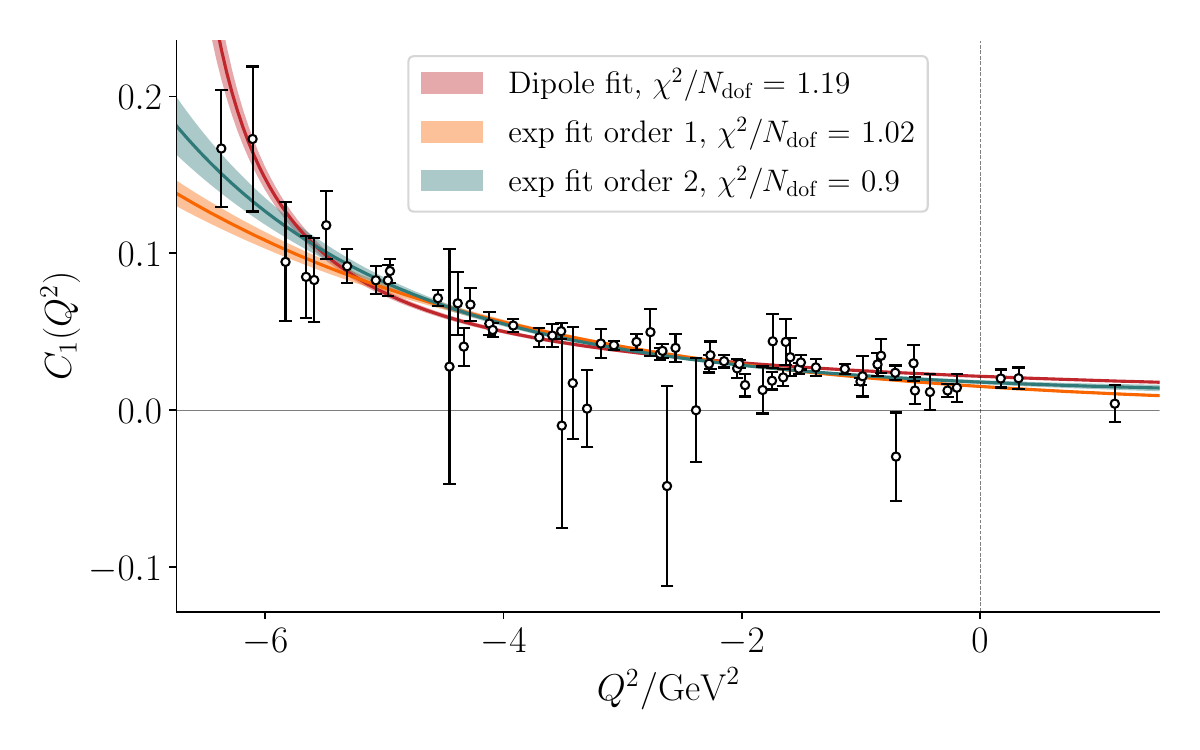}} 
\caption[]{\label{fig:eta-prime_Qsq_dipole_exp} Electric dipole (top) and longitudinal (bottom) form--factors for $h_c \to \gamma \eta'$ described by dipole and exponential fits. }
\end{figure}

\begin{figure}
  \centerline{\includegraphics[width=.5\textwidth]{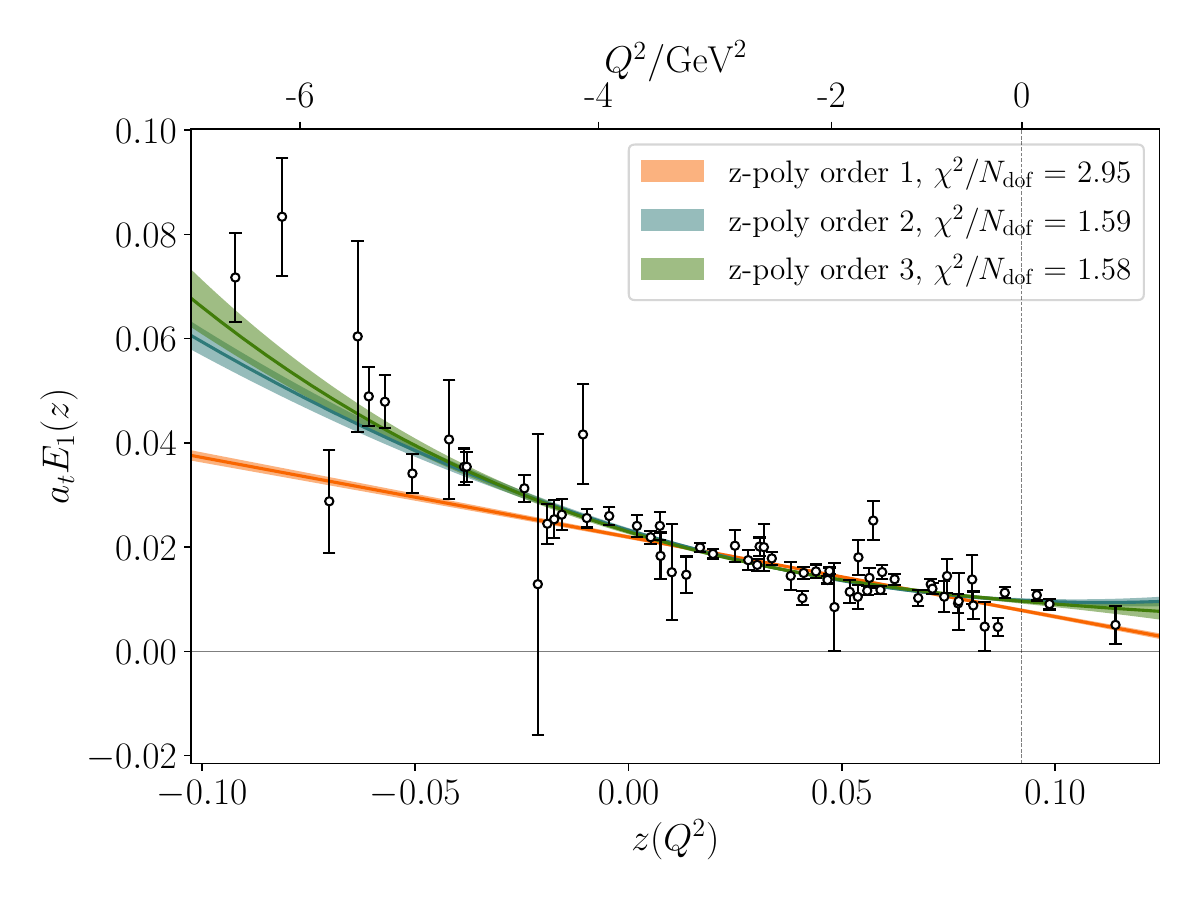}}
  \centerline{\includegraphics[width=.5\textwidth]{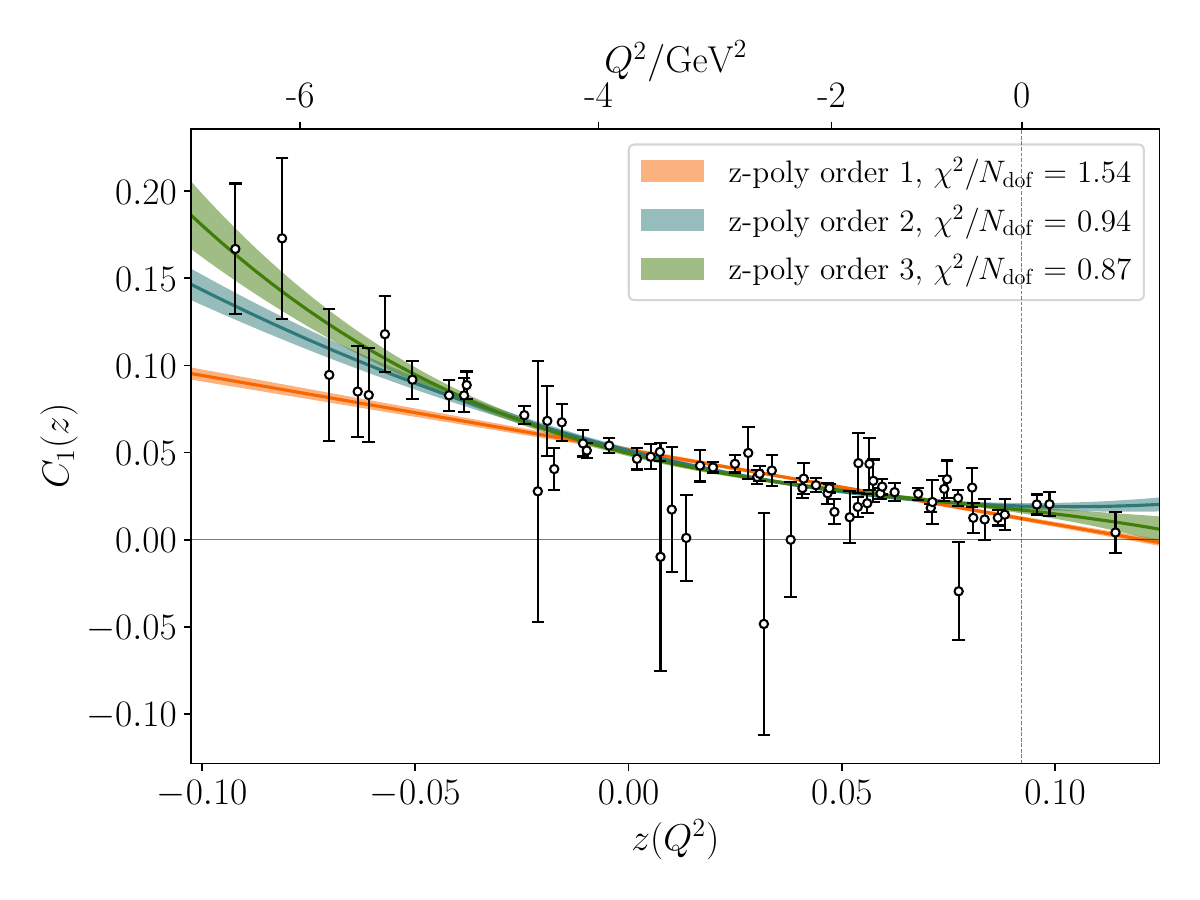}} 
\caption[]{\label{fig:eta-prime_Qsq_z_fits} Electric dipole (top) and longitudinal (bottom) form--factors for $h_c \to \gamma \eta'$ described using low--order polynomials in a conformally mapped variable $z(Q^2)$. }
\end{figure}

\pagebreak
Descriptions of the $Q^2$ dependence of these form--factors using the same forms used above are presented in Figures~\ref{fig:eta-prime_Qsq_dipole_exp} and \ref{fig:eta-prime_Qsq_z_fits}, where again we see generally good descriptions with the exception of the order 1 conformally mapped polynomial. The somewhat larger values of $\chi^2$ per d.o.f for fits which successfully capture the trend of the data reflect the larger degree of scatter in the data points.
The presence of more points in the region around $Q^2=0$ (due to the larger mass of the $\eta'$ relative to $\eta$) leads to a smaller parameterization dependence on $E_1(0)$. The fit parameter values for all presented fits are provided in Appendix~\ref{app3}.

\pagebreak\clearpage
\subsection{$\psi' \to \gamma \eta^{(\prime)}$}

\begin{figure}[h]
  \centerline{\includegraphics[width=.5\textwidth]{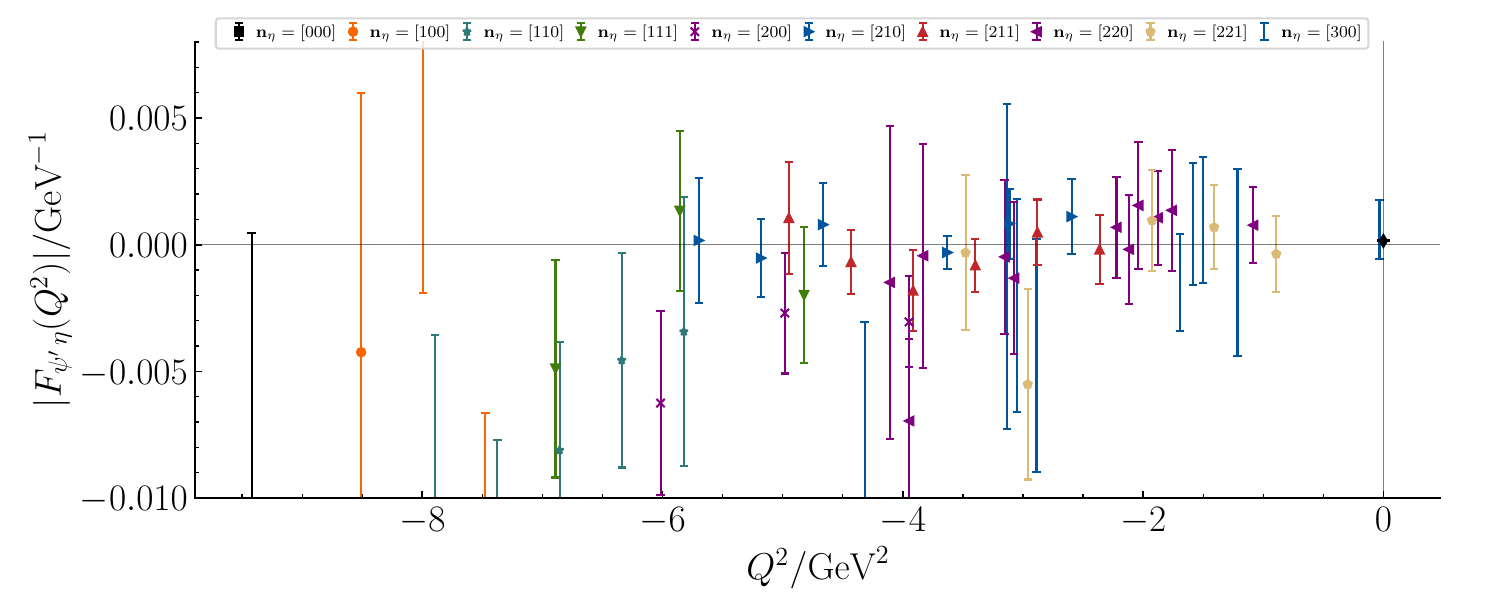}}
  \centerline{\includegraphics[width=.5\textwidth]{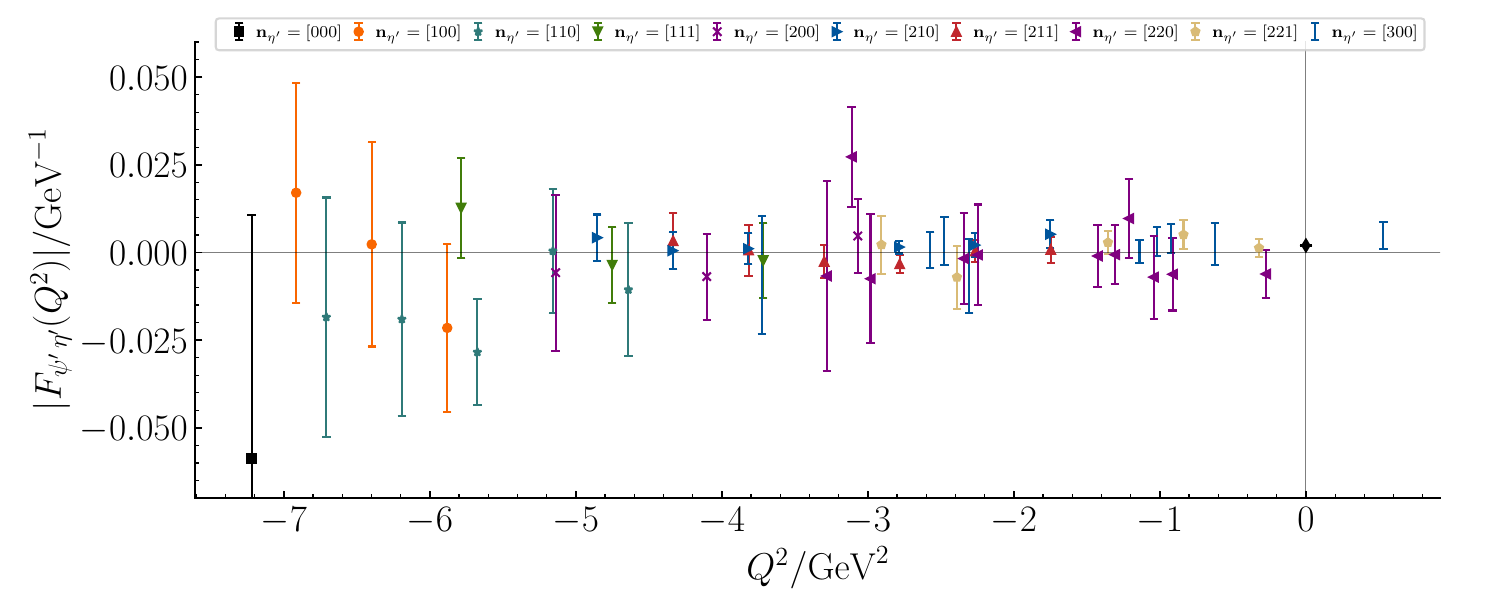}}
  \caption[]{\label{fig:eta-psiprime-allmom} Form--factor for $\psi' \to \gamma \eta$ (top) and $\psi' \to \gamma \eta'$ (bottom) as a function of photon virtuality, $Q^2$. These results extracted using a single source--sink separation, $\Delta t/a_t = 20$, with $\psi'$ momenta in the range $|\mathbf{n}_{\psi'}|^{2} \leq 4$. The black diamond indicates the value of the form--factor at $Q^{2}=0$ extracted from the experimental decay widths measured by BESIII~\cite{BESIII:2017nnc}.
  }
\end{figure}

As mentioned in the introduction, our optimized operator technology allows us to access the first excited vector charmonium state, the $\psi'$, and determine the form--factor for each of $\psi' \to \gamma \eta$ and $\psi' \to \gamma \eta'$. Here the presence of only a single form--factor simplifies the analysis, and our procedure largely matches that described in Ref.~\cite{Batelaan:2025vbb}. In Figure~\ref{fig:eta-psiprime-allmom} we show the extracted form--factor in the case of using just a single source--sink separation, $\Delta t /a_t = 20$, where we see that across the entire $Q^2$ range, the signal is broadly compatible with zero\footnote{A limited study using fewer $\psi'$ momenta, but more $\Delta t$ choices and simultaneous $\Delta t$ fitting, leads to quite similar results.}. Form--factor values for real photons corresponding to the measured BESIII decay widths~\cite{BESIII:2017nnc} are also shown, and we see that the level of statistical precision on our lattice results is not currently at a level where comparison to experiment is warranted.

\pagebreak
\section{Conclusions}\label{conclusions}

The previous sections show that the use of optimized operator technology for a large range of meson momenta can lead to internally consistent estimates of the ${h_c \to \gamma \eta^{(\prime)}}$ form--factors, $E_1(Q^2)$ and $C_1(Q^2)$, across the timelike region and slightly into the spacelike region. These estimates, sampled discretely in virtuality, $Q^2$, can be well-described by simple parameterizations that span the entire determined $Q^2$ region.

\begin{figure}
  \centerline{\includegraphics[width=.48\textwidth]{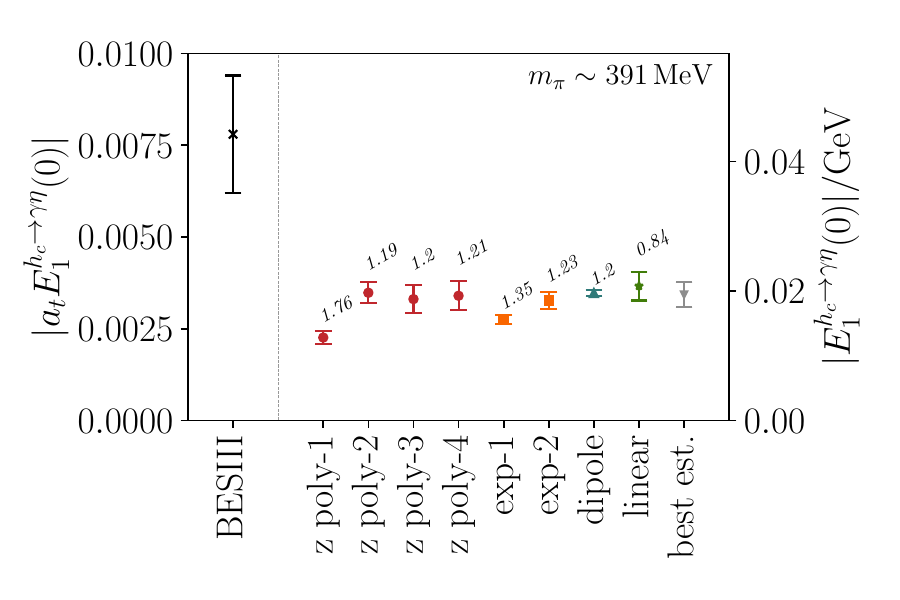}}
  \centerline{\includegraphics[width=.48\textwidth]{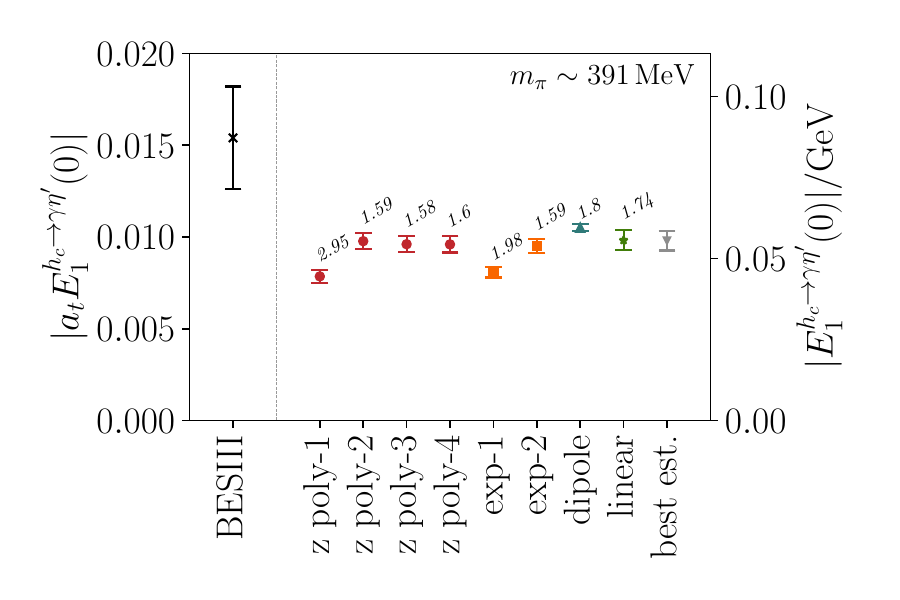}}
\caption[]{\label{fig:exp-comp} Electric dipole form--factor at $Q^2=0$ for $h_c \to \gamma \eta$ (top) and $h_c \to \gamma \eta'$ (bottom). Estimates following from various parameterizations of $Q^2$--dependence described in the text (labelled by their $\chi^2/{N_\mathrm{dof}}$ values) are compared to values extracted from the BESIII experimental decay widths~\cite{BESIII:2024xfs}.}
\end{figure}

The radiative decay rates for $h_c \to \gamma \eta$ and $h_c \to \gamma \eta'$ are controlled by $E_1(0)$, as presented in Eqn.~\ref{eqn:rad_decay}. Each parameterization of the $Q^2$--dependence of $E_1(Q^2)$, described in the previous section, yields an estimate of $E_1(0)$, to which we may add a ``more--local'' estimate coming from a simple linear interpolation between the handful of discrete points around $Q^2=0$. These various estimates are shown in Figure~\ref{fig:exp-comp} along with the values for $h_c \to \gamma \eta$ and $h_c \to \gamma \eta'$ extracted from experimental decay data.

\pagebreak
Considering the variation over parameterization choice, and variation under adjustment of the lattice anisotropy within a conservative error estimate (which is a modest effect), we give our best estimates of,
\begin{align*}
\big| E_1^{h_c \to \gamma \eta}(0) \big|  &= 0.0195(19)\, \mathrm{GeV} \, ,\\
\big| E_1^{h_c \to \gamma \eta'}(0) \big| &= 0.0556(30)\, \mathrm{GeV} \, .
\end{align*}

\bigskip
It is clear from Figure~\ref{fig:exp-comp} that, as was observed for $J/\psi \to \gamma \eta^{(\prime)}$ in Ref.~\cite{Batelaan:2025vbb}, the calculated $h_c \to \gamma \eta^{(\prime)}$ form--factor values for real photons lie significantly below the experimental values. This might indicate that the common feature of the two reactions, production of the $\eta$ and $\eta'$ by a gluonic intermediate state, is the cause of both discrepancies.

The computed $\eta'/\eta$ ratio, 
\begin{equation*}
\left| \frac{ E_1^{h_c \to \gamma \eta'}(0) }{ E_1^{h_c \to \gamma \eta}(0) } \right| =  2.9(3) \, ,
\end{equation*}
can be compared to the corresponding ratio for $J/\psi$ decays computed on the same lattices,
\begin{equation*}
\left|\frac{ F^{J/\psi \to \gamma \eta'}(0)  }{ F^{J/\psi \to \gamma \eta}(0)}  \right| = 3.3(3) \, , 
\end{equation*}
which we observe to be compatible within errors\footnote{and even more compatible for, $$
\left|\frac{ |\mathbf{q}_{\eta'}| F^{J/\psi \to \gamma \eta'}(0)  }{ |\mathbf{q}_{\eta}| F^{J/\psi \to \gamma \eta}(0)}  \right| = 3.1(3)\, ,$$ if one chooses to consider a factor of the transition momentum in the form--factor definition.}. That $\eta'$ production is larger than $\eta$ production is of course expected in the usual understanding of this system in which the $\eta'$ is dominated by an $SU(3)$ flavour \emph{singlet} component, with only a small admixture of \emph{octet}, while the $\eta$ is dominated by the \emph{octet}. Since only the \emph{singlet} can be reached in a purely gluonic intermediate state, the hierarchy is natural.

A dimensionless ratio that compares the strength of the \emph{electric dipole} process in $h_c$ decays to the \emph{magnetic dipole} process in $J/\psi$ decays, removing the differing phase--space, is given by $\frac{|E_1(0)|}{ m_{J/\psi} |\mathbf{q}| |F(0)|}$. Comparing the computed values on these lattices to the experimental values,
\begin{align*}
\frac{ |E_1(0)|}{ m_{J/\psi} |\mathbf{q}| |F(0)|}&\bigg|_{\eta} = 1.84(37)^\mathrm{expt.}\,,  1.76(20)^{391 \, \mathrm{MeV}} \\
\frac{ |E_1(0)|}{ m_{J/\psi} |\mathbf{q}| |F(0)|}&\bigg|_{\eta'} = 1.58(29)^\mathrm{expt.}\,,  1.63(11)^{391 \, \mathrm{MeV}} \, ,
\end{align*}
we see approximate agreement for both the $\eta$ and the $\eta'$. To the extent that the charmonium annihilation process factorizes from the production (from gluons) of the $\eta$ or $\eta'$, this is what we would expect if the origin of the discrepancy lies with properties of the $\eta$ and $\eta'$ on these lattices.

\begin{figure}[b]
  \centerline{\includegraphics[width=.5\textwidth]{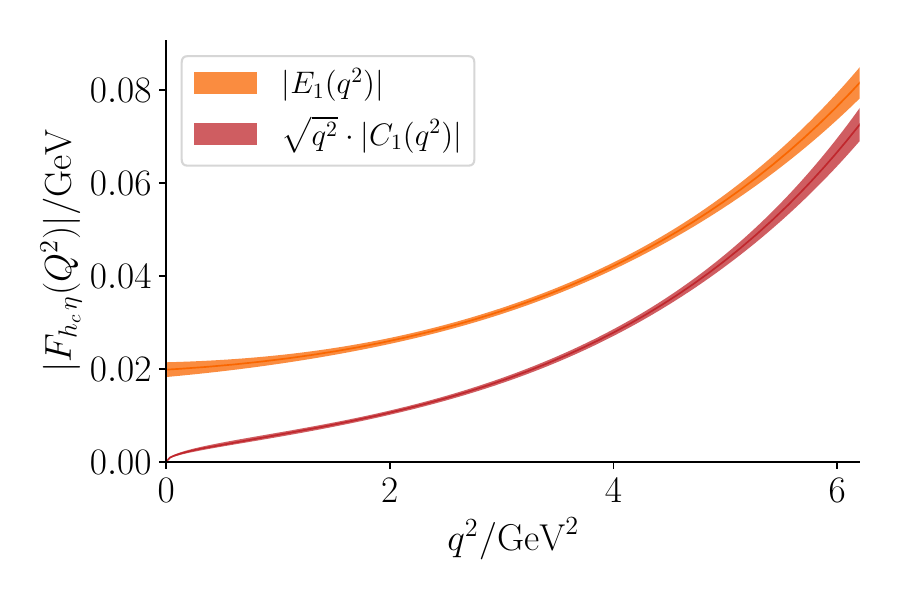}}
  \centerline{\includegraphics[width=.5\textwidth]{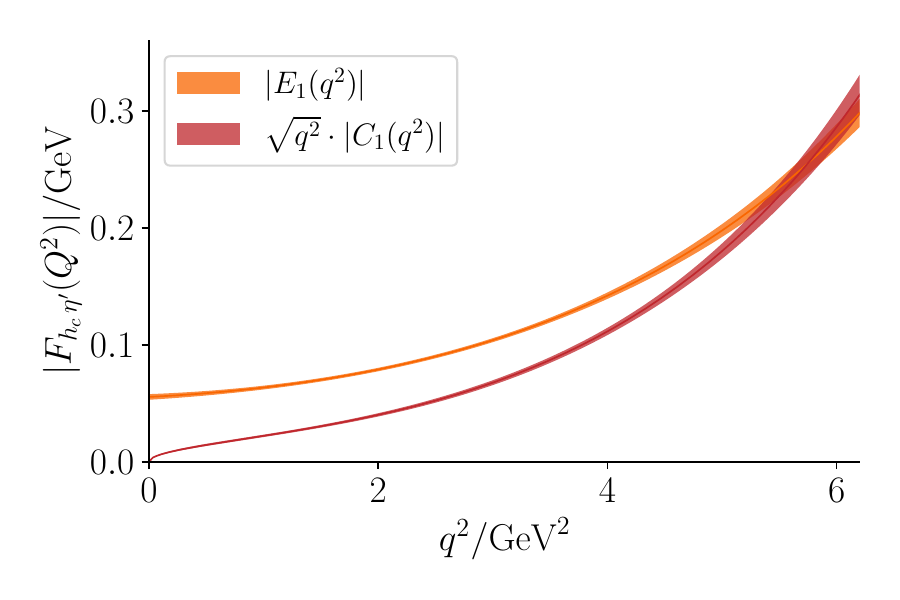}}
  \caption[]{\label{fig:Dalitz-fits} 
  Form-factors, $E_1(q^2)$ and $\sqrt{q^2} C_1(q^2)$, for $h_c \to \gamma \eta$ (top) and $h_c \to \gamma \eta'$ (bottom), across the timelike region of photon virtuality, as relevant for Dalitz decays of the $h_c$. Parameterized description is by an order--2 polynomial in the conformally--mapped variable, $z(Q^2)$. 
  }
\end{figure}

\bigskip 
Dalitz decays, $h_c \to \ell^+ \ell^- \, \eta^{(\prime)}$, depend upon both the electric dipole and longitudinal form--factors for timelike $q^2 = -Q^2$ from $(2m_\ell)^2$ to $(m_{h_c} - m_{\eta^{(\prime)}})^2$. Our analysis does not quite extend to the maximum kinematically allowed $q^2$ value, but the trend can be observed. 
Motivated by the particular combinations of $E_1$ and $C_1$ appearing in Eqn.~\ref{eqn:dalitz}, in Figure~\ref{fig:Dalitz-fits} we plot $E_1(q^2)$ and $\sqrt{q^2} \, C_1(q^2)$ across the timelike region using the description of our data given by an order--2 polynomial in the conformally mapped variable, $z$. 
We observe that the electric dipole gives the larger contribution to the decay to $\eta$ across the entire considered range, but that in general both form--factors will be significant.

\pagebreak

We conclude by restating that the most likely origin of the observed discrepancy with respect to experiment common to both $J/\psi \to \gamma \eta^{(\prime)}$ and $h_c \to \gamma \eta^{(\prime)}$ is some property of the $\eta$ and $\eta'$ on this particular lattice. The importance of the SU(3) flavor singlet content of these mesons in this process, and the sensitivity of that component to the $U(1)$ axial anomaly, motivates consideration on other lattices where the gauge--field topology sampling is better studied. The results of this paper and Ref.~\cite{Batelaan:2025vbb} show that relevant signals with a high degree of statistical precision can be obtained by use of contemporary techniques. The particular sensitivity of the $\eta, \eta'$ system to detailed properties of the gauge--field configurations may not be shared by other light--meson final states. 

Returning to the use of charmonium radiative decays as a light meson resonance factory, a next step will be to consider multi--hadron production, such as $J/\psi \to \gamma \, \pi \pi$ and the coupled--channel $J/\psi \to \gamma \, K \overline{K}$, in which examples of scalar and tensor isoscalar resonances appear~\cite{BESIII:2015rug, BESIII:2018ubj, Rodas:2021tyb, Sarantsev:2021ein, Briceno:2017qmb}. These processes will be accessible in lattice calculations by combining the technology developments presented in this paper and in Ref.~\cite{Batelaan:2025vbb}, with the finite--volume formalism of Refs.~\cite{Lellouch:2000pv,Briceno:2014uqa, Briceno:2015csa, Briceno:2021xlc}, as demonstrated (in other channels) in Refs.~\cite{Briceno:2015dca,Briceno:2016kkp,Andersen:2018mau,Alexandrou:2018jbt,Radhakrishnan:2022ubg,Ortega-Gama:2024rqx, Leskovec:2025gsw}.

\begin{acknowledgments}

  We thank our colleagues within the Hadron Spectrum Collaboration for their continued assistance. The authors acknowledge support from the U.S. Department of Energy contract DE-SC0018416 at William \& Mary, and contract DE-AC05-06OR23177, under which Jefferson Science Associates, LLC, manages and operates Jefferson Lab.
This material is based upon work supported by the U.S. Department of Energy, Office of Science, Office of Nuclear Physics under Contract No. 89243126CSC000213.
This work contributes to the goals of the U.S. Department of Energy \emph{ExoHad} Topical Collaboration, Contract No. DE-SC0023598.
The authors acknowledge support from the U.S. Department of Energy, Office of Science, Office of Advanced Scientific Computing Research and Office of Nuclear Physics, Scientific Discovery through Advanced Computing (SciDAC) program. 
Also acknowledged is support from the Exascale Computing Project (17-SC-20-SC), a collaborative effort of the U.S. Department of Energy Office of Science and the National Nuclear Security Administration.
MB is supported by the Australian Research Council grants DP240102839 and DP260104445.

This work used clusters at Jefferson Laboratory under the USQCD Initiative and the LQCD ARRA project. This work also used the Cambridge Service for Data Driven Discovery (CSD3), part of which is operated by the University of Cambridge Research Computing Service (www.csd3.cam.ac.uk) on behalf of the STFC DiRAC HPC Facility (www.dirac.ac.uk). The DiRAC component of CSD3 was funded by BEIS capital funding via STFC capital grants ST/P002307/1 and ST/R002452/1 and STFC operations grant ST/R00689X/1. Other components were provided by Dell EMC and Intel using Tier-2 funding from the Engineering and Physical Sciences Research Council (capital grant EP/P020259/1). 
DiRAC is part of the National E-Infrastructure.

Also used was an award of computer time provided by the U.S.\ Department of Energy INCITE program and supported in part under an ALCC award, and resources at: the Oak Ridge Leadership Computing Facility, which is a DOE Office of Science User Facility supported under Contract DE-AC05-00OR22725; the National Energy Research Scientific Computing Center (NERSC), a U.S.\ Department of Energy Office of Science User Facility located at Lawrence Berkeley National Laboratory, operated under Contract No. DE-AC02-05CH11231; the Texas Advanced Computing Center (TACC) at The University of Texas at Austin; the Extreme Science and Engineering Discovery Environment (XSEDE), which is supported by National Science Foundation Grant No. ACI-1548562; and part of the Blue Waters sustained-petascale computing project, which is supported by the National Science Foundation (awards OCI-0725070 and ACI-1238993) and the state of Illinois. Blue Waters is a joint effort of the University of Illinois at Urbana-Champaign and its National Center for Supercomputing Applications.

The software codes
{\tt Chroma}~\cite{Edwards:2004sx}, {\tt QUDA}~\cite{Clark:2009wm,Babich:2010mu}, {\tt QUDA-MG}~\cite{Clark:SC2016}, {\tt QPhiX}~\cite{ISC13Phi},
{\tt MG\_PROTO}~\cite{MGProtoDownload}, and {\tt QOPQDP}~\cite{Osborn:2010mb,Babich:2010qb} were used. 

\end{acknowledgments}

\bibliography{Library}

\appendix
\widetext

\section{Consistency between ground state and excited state $h_c$ projected operators}\label{app1}

As discussed in Section~\ref{two-point}, the $h_c$ appears as the ground--state only in irreps containing helicity 0, namely $[000]\,T_1$ and $\mathbf{n} A_2$, while in irreps $\mathbf{n}B_1$, $\mathbf{n} B_2$, and $\mathbf{n} E_2$, the $h_c$ appears as an excited state above the ground--state $J/\psi$. In our main analysis we assume that the optimized operator technology works as expected and isolates the $h_c$ in both cases, and we combine these different irreps together in the linear algebra solution yielding $E_1$ and $C_1$ at each $Q^2$.

\begin{figure*}
  \centerline{\includegraphics[width=.8\textwidth]{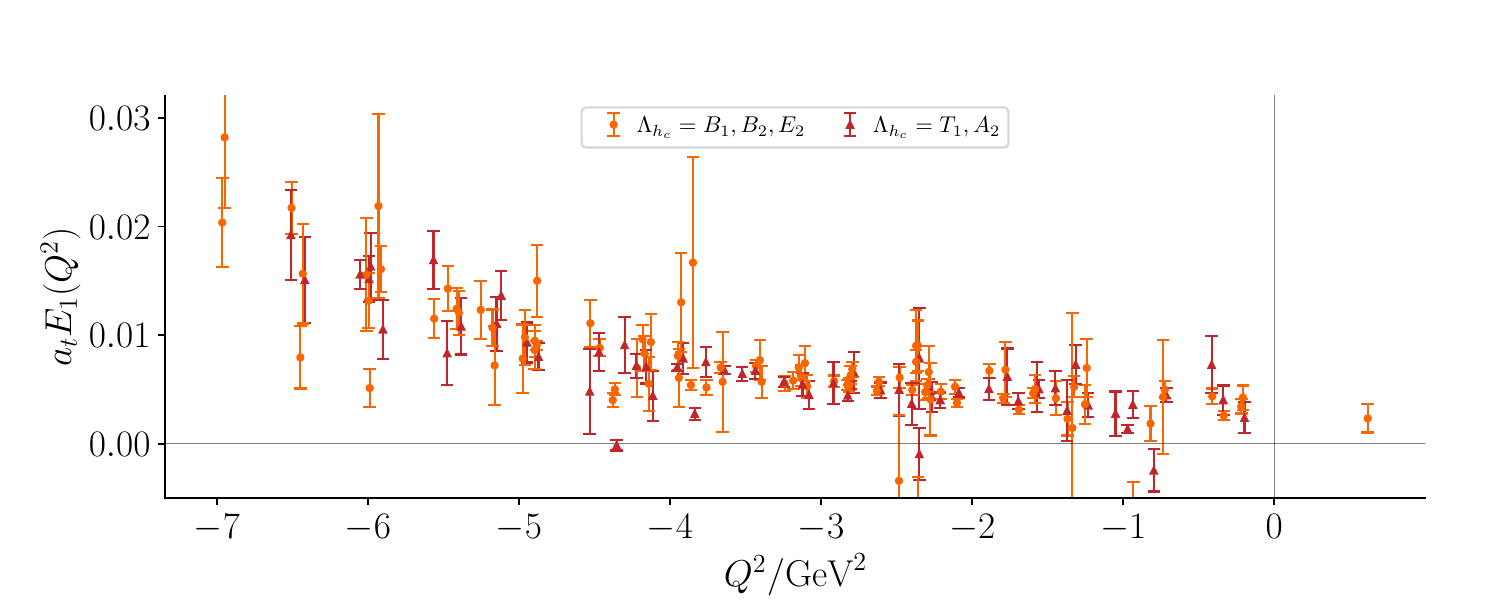}}
  \centerline{\includegraphics[width=.8\textwidth]{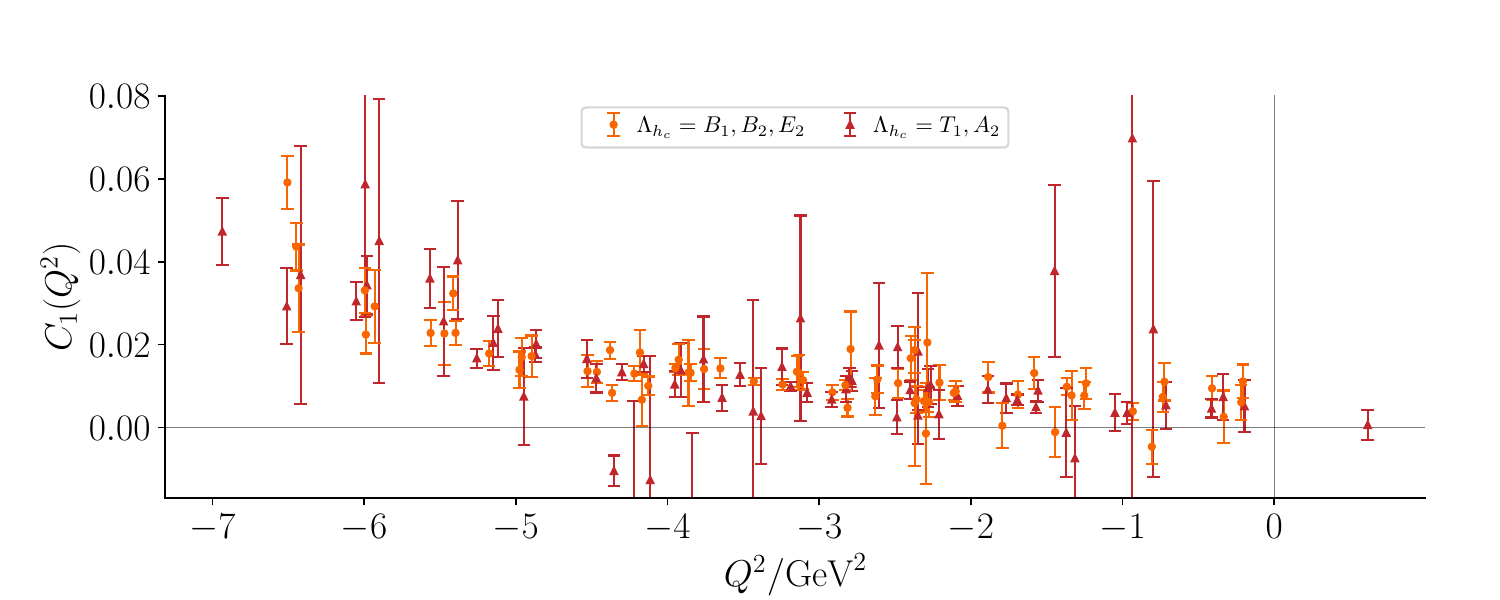}}
  \caption[]{\label{fig:eta-hc-irreps} Electric dipole (top) and longitudinal (bottom) form-factors for $h_c \to \gamma \eta$ as a function of photon virtuality, $Q^2$. Red triangles indicate points for which the $h_c$ is the ground--state in its irrep, and orange circles indicate points for which the $h_c$ is an excited state in its irrep.
  }
\end{figure*}

We can also choose not to do this, and instead treat the cases in which the $h_c$ is an excited state separately from those where it is the ground--state. Figure \ref{fig:eta-hc-irreps} shows the resulting form--factors for the $h_c \to \gamma \eta$ case, where we observe compatible estimates from the two cases. We argue that this demonstrates both the practical success of the use of optimized operators, and also the lack of any significant discretization errors impacting the subduction of the Lorentz covariant decomposition into different irreps.

\section{Simultaneous fitting of correlation functions with different $\Delta t$.}\label{app4}

\begin{figure*}
  \centerline{\includegraphics[width=.8\textwidth]{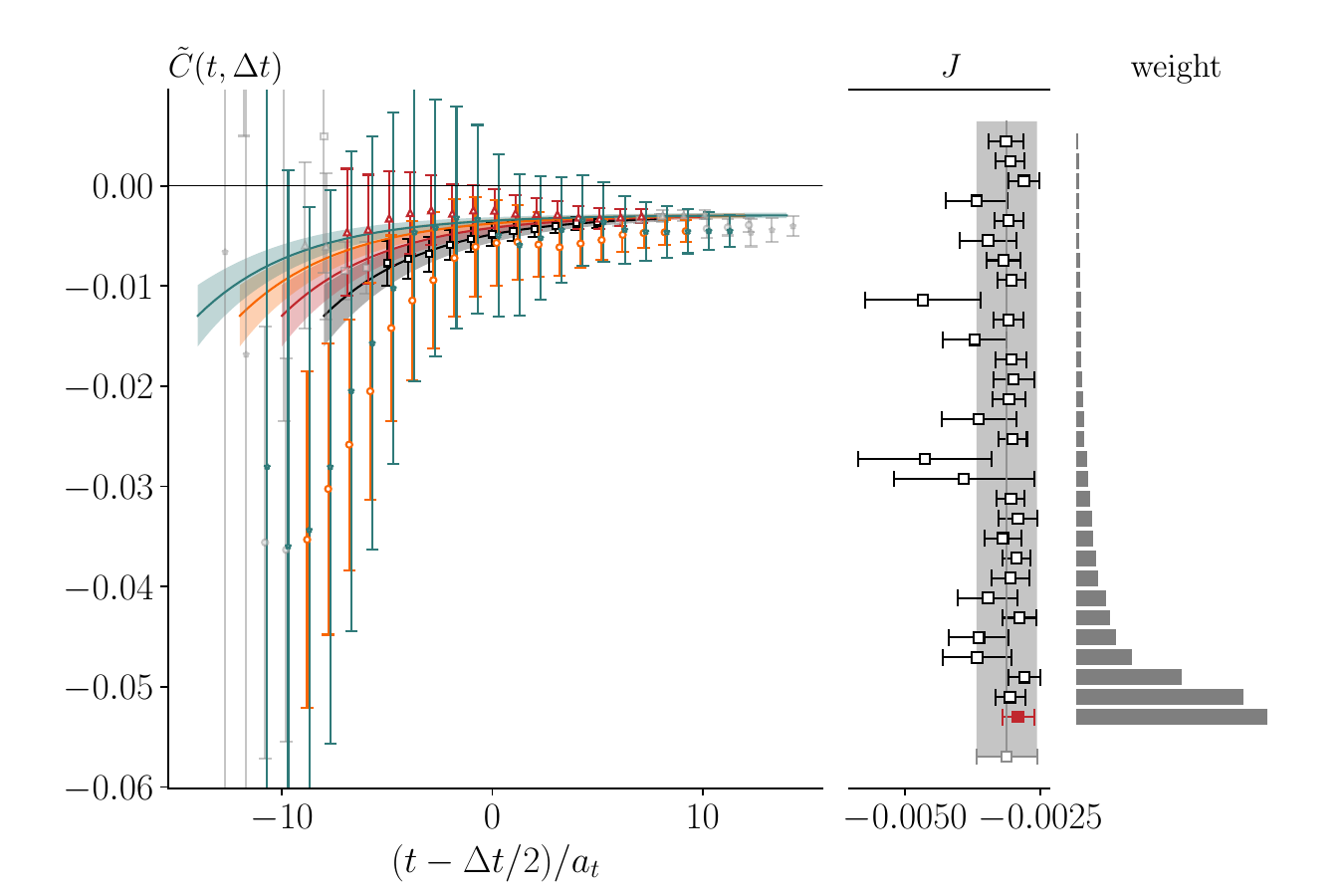}}
  \caption[]{\label{fig:AIC-ex} (left) Description of the same data as plotted in Figure~\ref{fig:proj0-dt-comp1}. (right) 30 highest AIC fits to this data, varying time--windows and fit--forms, together with the model--averaged final value of $J$. 
  }
\end{figure*}

Our procedure for fitting multiple correlators having the same operators for differing source--sink separations, $\Delta t$, was briefly described in the text. A large number of fits are considered, and Figure~\ref{fig:AIC-ex} illustrates how the model--averaging combines fits over differing time--windows and fit--forms into a single estimate for the value of $J$. The fit with the single highest value of AIC is again shown (as was shown in Figure~\ref{fig:proj0-dt-comp1}) alongside the extracted values of $J$ for the 30 fits with the highest AIC values. Treating the AIC values as probability weights, these estimates can be used in a weighted average to yield the final determination of $J$, which is observed to reflect the preference in some (lower probability) fits for a slightly lower mean value than in the overall highest probability fits.

\begin{figure*}
  \centerline{\includegraphics[width=.98\textwidth]{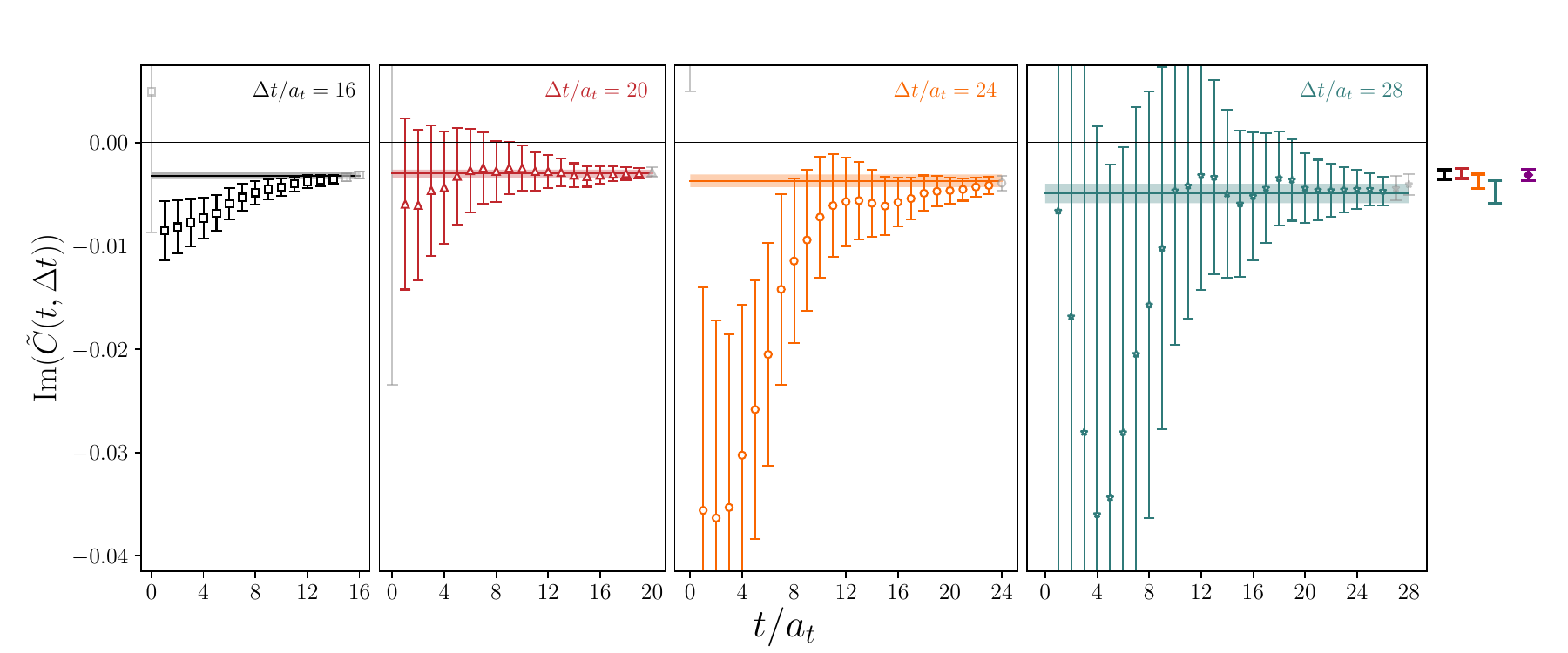}}
  \caption[]{\label{fig:fit-comp} Correlators previously shown in Figure \ref{fig:proj0-dt-comp1}. Panels show the highest AIC--value fit to each $\Delta t$-correlator separately, with the points on the right side showing the model--average $J$ value, taken over 30 variations in time--window and fit--form. These separate $\Delta t$ determinations are compared with each other, and with the result of simultaneous fitting (purple diamond), as used in the text. 
  }
\end{figure*}

\smallskip
The approach used in this paper can be compared to the method used in Ref.~\cite{Batelaan:2025vbb} of fitting each $\Delta t $ independently, and checking for approximate consistency in the estimates.
In Figure \ref{fig:fit-comp} we show the result of performing independent fits to each $\Delta t$, allowing for the same kinds of variations of Eqn.~\ref{eqn:tslice_fit} over many choices of time--window. For each $\Delta t$, the best single timeslice description, as measured by AIC--value is plotted, along with the model--averaged estimate of $J$. We observe that the independent fits to each $\Delta t$ are broadly consistent with each other, and with the simultaneous fitting approach.

\section{Retaining only reliable extractions of the $E_1$, $C_1$ system.}\label{app2}

\begin{figure*}
\includegraphics[width=.6\textwidth]{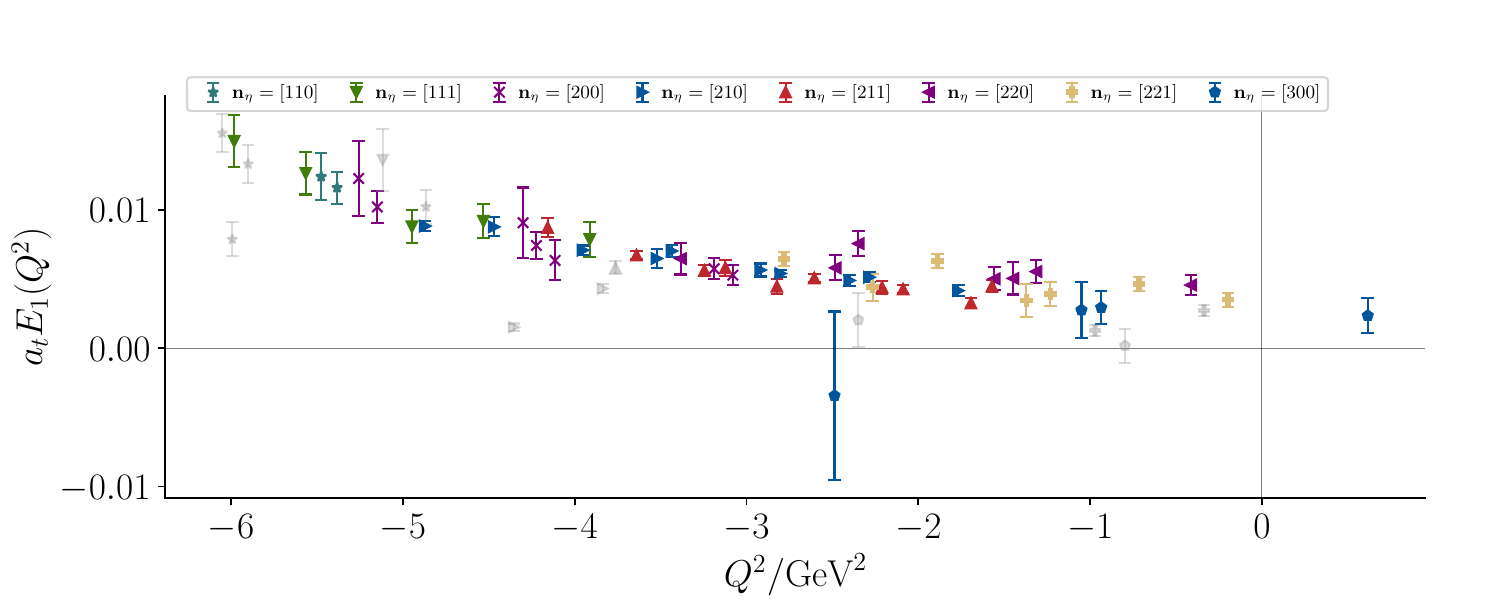}
\includegraphics[width=.6\textwidth]{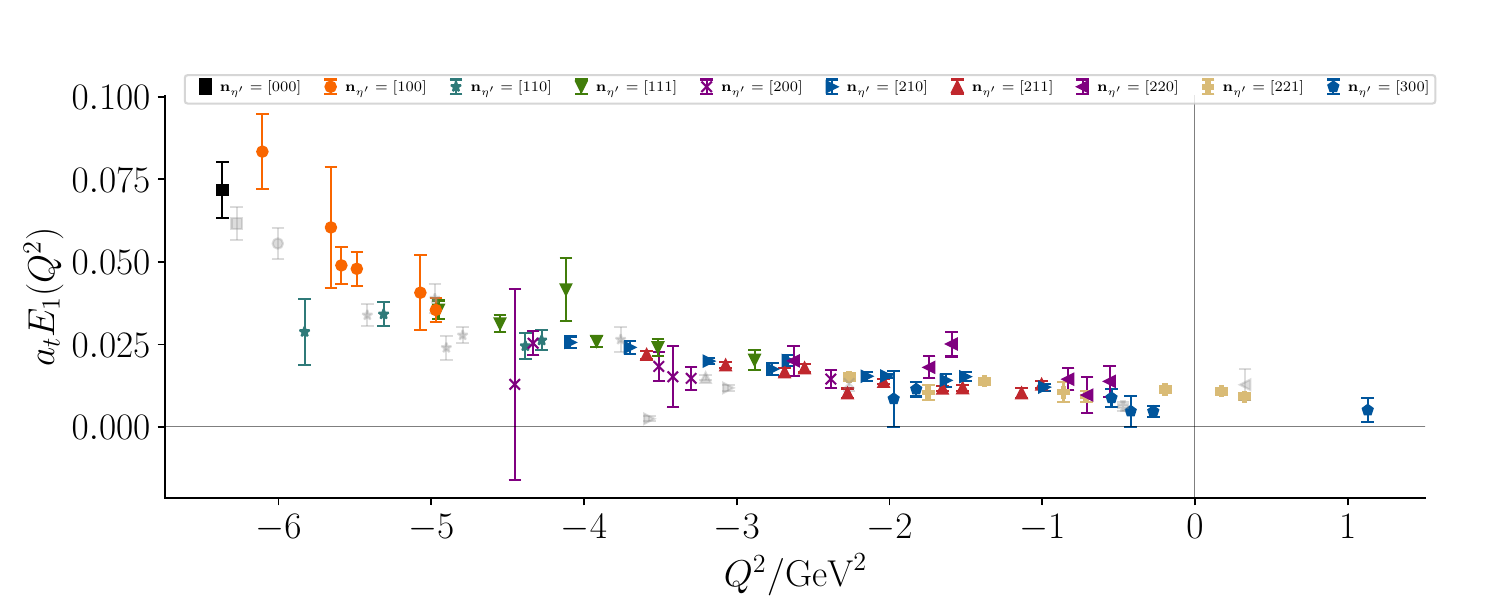}
\vspace*{-3mm}
  \caption[]{\label{fig:eta-chisq} Electric dipole (top) and longitudinal (bottom) form-factors for $h_c \to \gamma \eta$ as a function of photon virtuality, $Q^2$.
Grayed-out points correspond to solution by Eqn.~\ref{eqn:SVD} corresponding to $\chi^2$ per degree of freedom greater than 2.0. These points are excluded from the figures in the main text and are not used further in analysis.
  }
\end{figure*}

\begin{figure*}
\includegraphics[width=.6\textwidth]{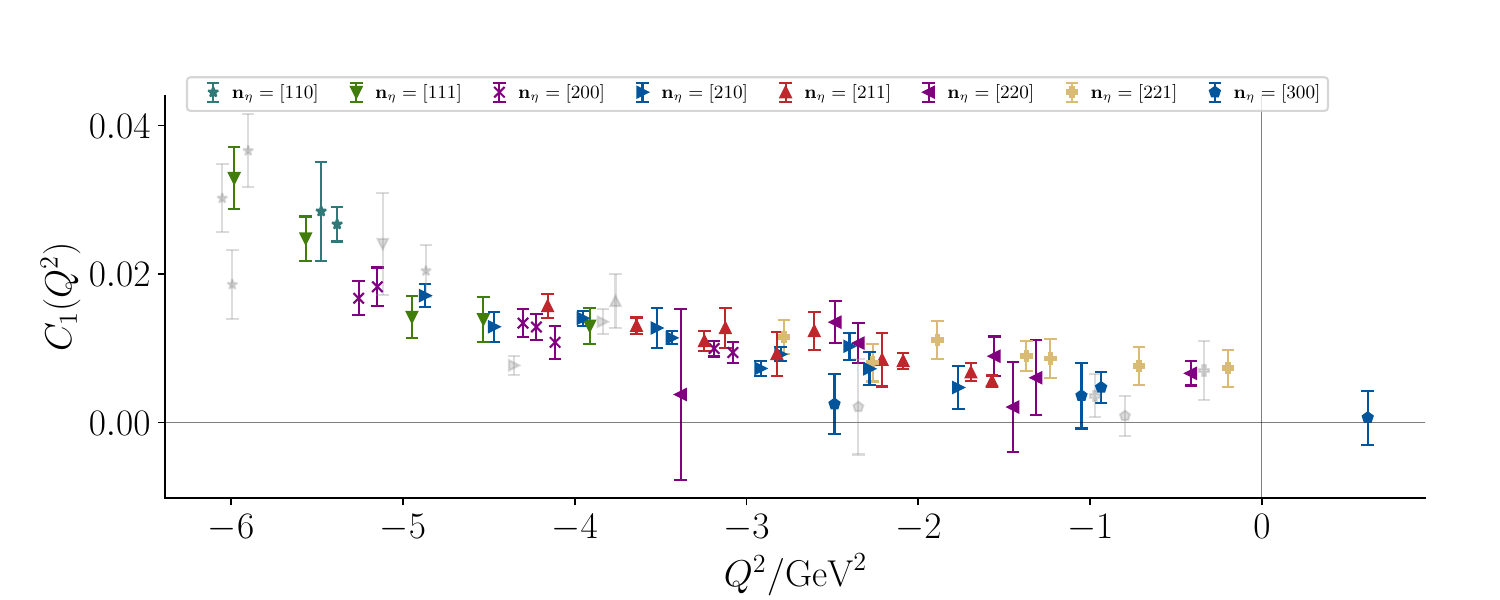}
\includegraphics[width=.6\textwidth]{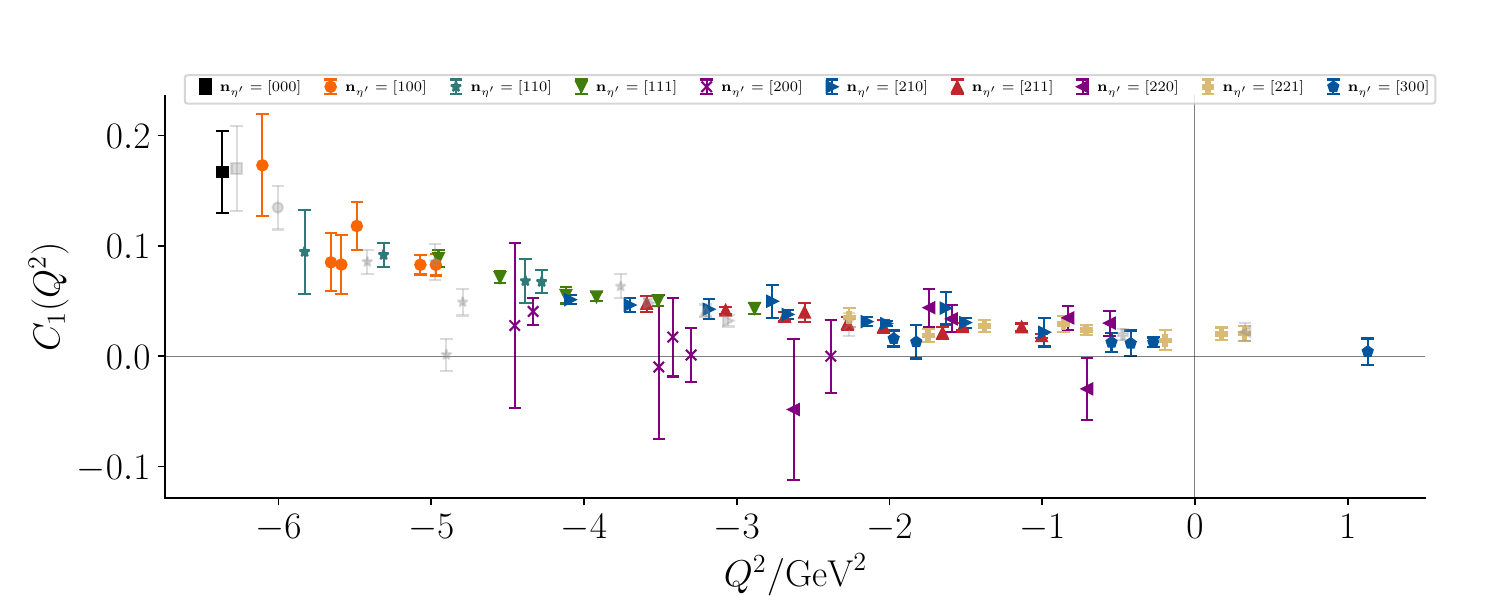}
\vspace*{-3mm}
  \caption[]{\label{fig:eta-chisqprime} As Figure~\ref{fig:eta-chisq} but for $h_c \to \gamma \eta'$.
  }
\end{figure*}

As described in Section~\ref{three-point}, in order to be conservative, we choose to eliminate from further analysis any $Q^2$ points where the description by Eqn.~\ref{eqn:SVD} of the set of matrix--elements corresponds to a large value of $\chi^2$ per degree of freedom. Applying a cut will help to remove any instances where imperfect timeslice fitting leads to incompatible matrix--element values appearing in $\mathbf{\Gamma}$. In Figures~\ref{fig:eta-chisq}, \ref{fig:eta-chisqprime}  we show both form--factors for $h_c \to \gamma \eta^{(\prime)}$, with the grayed--out points being those excluded because they have $\chi^2_\Gamma / (N-2) > 2.0$.

\clearpage
We observe that in many cases this cutoff may be overly cautious as the extracted $E_1$, $C_1$ values are in good agreement with the retained data points, but we do observe some outlier points, and points with suspiciously small statistical errors which might overly influence fits to the $Q^2$ dependence of the form-factors were they to be retained in analysis.

\section{Description of the $Q^2$ dependence of form--factors.}\label{app3}

In Section~\ref{form_factors} we presented the description of our discrete data on $E_1(Q^2)$ and $C_1(Q^2)$ in terms of parameterizations previously used to describe the single form-factor for $J/\psi \to \gamma \eta^{(\prime)}$ in Ref.~\cite{Batelaan:2025vbb}. In this appendix we provide further details of these fits.
With the exception of the linear interpolation between a few points straddling $Q^2=0$, the fits use the whole data set for each of $h_c \to \gamma \eta$ and $h_c \to \gamma \eta'$. The data points have correlations that are modest with the eigenvalues of the covariance matrix in each case satisfying $\lambda_{\text{min}}/\lambda_{\text{max}} \approx 0.07$.

\smallskip
\noindent The ``dipole'' form referred to in the text is
\begin{equation*}
\label{eq:11}
F(Q^2) = \frac{F_0}{1+Q^2/\Lambda^2}\, ,
\end{equation*}
which clearly has a pole singularity at $Q^2 = -\Lambda^2$ which, given that the data shows no divergence, must lie outside the $Q^2$ range we consider. The fit parameters are:

\medskip
\centerline{
 \renewcommand{\arraystretch}{1.4}
 \begin{tabular}{r @{\hspace{5mm}} rr @{\hspace{5mm}} rr}
   & $E_{1}^{h_c \to \gamma\eta}$ & $C_{1}^{h_c \to \gamma\eta}$ & $E_1^{h_{c} \to \gamma\eta'}$ & $C_1^{h_{c} \to \gamma\eta'}$\\
   \hline
   $F_0/\textrm{GeV}$  & 0.01969(52) & 0.0334(12) & 0.0595(10) & 0.1223(34)  \\
   $\Lambda/\textrm{GeV}$  & 2.783(25) & 2.692(22) & 2.705(13) & 2.651(13) \\
   \hline
   $\chi^2/N_{\textrm{dof}}$ & 1.2 & 1.24 & 1.8 & 1.19 \\
 \end{tabular}
}
\smallskip

\bigskip
\noindent The ``exponential'' forms referred to in the text are
\begin{align*}
  F(Q^2) &= F_0\, e^{-Q^2/16\beta^2}, \\
  F(Q^2) &= F_0\, e^{-(1+\alpha Q^2)\, Q^2/16\beta^2} \, ,
\end{align*}
which have no nearby singularities by construction. The fit parameters are:

\medskip
\centerline{
 \renewcommand{\arraystretch}{1.4}
 \begin{tabular}{r @{\hspace{5mm}} rr @{\hspace{5mm}} rr}
   & $E_{1}^{h_c \to \gamma\eta}$ & $C_{1}^{h_c \to \gamma\eta}$ & $E_1^{h_{c} \to \gamma\eta'}$ & $C_1^{h_{c} \to \gamma\eta'}$\\
   \hline
   $F_0/\textrm{GeV}$  & 0.01559(71) & 0.0221(16) & 0.0458(17)  & 0.0857(49) \\
   $\beta/\textrm{GeV}$  & 0.499(11)  & 0.433(11) & 0.4734(96) & 0.437(10) \\
   \hline
   $\chi^2/N_{\textrm{dof}}$ & 1.35 & 1.19 & 1.98 & 1.02 \\
 \end{tabular}
}

\bigskip 

\centerline{
 \renewcommand{\arraystretch}{1.4}
 \begin{tabular}{r @{\hspace{5mm}} rr @{\hspace{5mm}} rr}
   & $E_{1}^{h_c \to \gamma\eta}$ & $C_{1}^{h_c \to \gamma\eta}$ & $E_1^{h_{c} \to \gamma\eta'}$ & $C_1^{h_{c} \to \gamma\eta'}$\\
   \hline
   $F_0/\textrm{GeV}$  & 0.0185(13) & 0.0251(28) & 0.0539(23) &  0.1014(72)\\
   $ \beta/\textrm{GeV}$  & 0.68(11)  & 0.506(60)  & 0.647(55) & 0.574(67) \\
   $ \alpha/\textrm{GeV}^{-2}$  & $-$0.126(80) & $-$0.042(42) & $-$0.150(52) & $-$0.120(65) \\
   \hline
   $\chi^2/N_{\textrm{dof}}$ & 1.23 & 1.19 & 1.59 & 0.9 \\
 \end{tabular}
}
\smallskip

\bigskip
\noindent Mapping $Q^2$ into a variable,
\begin{equation*}
z(Q^2) = \frac{\sqrt{t_{\text{cut}} + Q^2} - \sqrt{t_{\text{cut}} - t_{0}}}{\sqrt{t_{\text{cut}} + Q^2} + \sqrt{t_{\text{cut}} - t_{0}}} \, ,
\end{equation*}
with $\sqrt{t_\mathrm{cut}} = 3.490\, \mathrm{GeV}$ and $(\sqrt{t_0})_\eta = 1.868\, \mathrm{GeV}$, $(\sqrt{t_0})_{\eta'} = 1.941\, \mathrm{GeV}$, we considered polynomial descriptions of the data,
\begin{equation*}
F(Q^2) = \sum_{n=0}^k a_n \, z(Q^2)^n \, ,
\end{equation*}
up to cubic order. The fit parameters are:

\smallskip
\centerline{
 \renewcommand{\arraystretch}{1.4}
 \begin{tabular}{r @{\hspace{5mm}} rr @{\hspace{5mm}} rr}
   & $E_{1}^{h_c \to \gamma\eta}$ & $C_{1}^{h_c \to \gamma\eta}$ & $E_1^{h_{c} \to \gamma\eta'}$ & $C_1^{h_{c} \to \gamma\eta'}$\\
   \hline
   $a_0/\textrm{GeV}$ & 0.03798(64)  & 0.0700(18)   & 0.1243(22)   & 0.2918(80) \\
   $a_1/\textrm{GeV}$ & $-$0.294(14) & $-$0.687(36) & $-$0.865(36) & $-$2.42(13) \\
   \hline
   $\chi^2/N_{\textrm{dof}}$ & 1.76 & 1.57 & 2.95 &  1.54\\[0.5ex]
   $|F(0)|/\textrm{GeV}$ & 0.0128(11) & 0.0112(29) & 0.0445(21) & 0.0687(72) \\
 \end{tabular}
}

\bigskip

\centerline{
 \renewcommand{\arraystretch}{1.4}
 \begin{tabular}{r @{\hspace{5mm}} rr @{\hspace{5mm}} rr}
   & $E_{1}^{h_c \to \gamma\eta}$ & $C_{1}^{h_c \to \gamma\eta}$ & $E_1^{h_{c} \to \gamma\eta'}$ & $C_1^{h_{c} \to \gamma\eta'}$\\
   \hline
   $a_0/\textrm{GeV}$ & 0.03673(68)  & 0.0667(19)   & 0.1316(24)   & 0.2897(80) \\
   $a_1/\textrm{GeV}$ & $-$0.334(15) & $-$0.747(38) & $-$1.412(70) & $-$3.52(22) \\
   $a_2/\textrm{GeV}$ & 1.58(29) & 3.31(73)& 6.35(69) & 17.0(28) \\
   \hline
   $\chi^2/N_{\textrm{dof}}$ & 1.19 & 1.19 & 1.59 & 0.94 \\[0.5ex]
   $|F(0)|/\textrm{GeV}$ & 0.0197(16) & 0.0270(46) & 0.0554(24) & 0.1094(98) \\
 \end{tabular}
}

\bigskip

\centerline{
 \renewcommand{\arraystretch}{1.4}
 \begin{tabular}{r @{\hspace{5mm}} rr @{\hspace{5mm}} rr}
   & $E_{1}^{h_c \to \gamma\eta}$ & $C_{1}^{h_c \to \gamma\eta}$ & $E_1^{h_{c} \to \gamma\eta'}$ & $C_1^{h_{c} \to \gamma\eta'}$\\
   \hline
   $a_0/\textrm{GeV}$ & 0.03650(76)  & 0.0663(19)   & 0.1299(26)   & 0.2801(90) \\
   $a_1/\textrm{GeV}$ & $-$0.324(21) & $-$0.695(62) & $-$1.409(70) & $-$3.45(22) \\
   $a_2/\textrm{GeV}$ & 1.72(35)     & 3.74(83)     &  8.3(15)     & 27.3(52)\\
   $a_3/\textrm{GeV}$ & $-$4.2(60)   & $-$18.(17)   & $-$20.(14)   & $-$124.(54)\\
   \hline
   $\chi^2/N_{\textrm{dof}}$ & 1.2 & 1.19 & 1.58 & 0.87\\[0.5ex]
   $|F(0)|/\textrm{GeV}$ & 0.0187(22) & 0.0227(61) & 0.0544(25) & 0.097(11)\\
 \end{tabular}
}
\smallskip

\end{document}